# Electrochemical and thermal control of continuous phase transitions in P2-$Na_xNi_{1/3}Mn_{2/3}O_2$

Dylan A. Edelman[1], John Cattermull[1], Jue Liu[2], Zhelong Jiang[1,3], Hari Ramachandran[1], Edward Mu[4], Cheng Li[2], Anton Van der Ven[5], Katherine J. Harmon[1*], William C. Chueh[1,6,7*]

*[1]Department of Materials Science and Engineering, Stanford University, Stanford, CA, 94305, USA.*

*[2]Neutron Scattering Division, Oak Ridge National Laboratory, Oak Ridge, TN, 37831, USA.*

*[3]SLAC-Stanford Battery Center, Applied Energy Division, SLAC National Accelerator Laboratory, Menlo Park, CA, 94025, USA.*

*[4]Department of Chemistry, Stanford University, Stanford, CA, 94305, USA.*

*[5]Materials Department, University of California, Santa Barbara, Santa Barbara, CA, 93117, USA.*

*[6]Stanford Institute for Materials and Energy Sciences, SLAC National Accelerator Laboratory, Menlo Park, CA, 94025, USA.*

*[7]Department of Energy Science and Engineering, Stanford University, Stanford, CA, 94305, USA.*

**Corresponding author e-mails: kharmon@stanford.edu, wchueh@stanford.edu*

## Abstract

Sodium layered oxides often undergo phase transformations involving ordering or disordering of $Na^+$ upon desodiation, i.e., when cycled as a battery electrode. Accurately characterizing these phases is crucial for understanding functional properties, such as chemical diffusivity. In this work, we reveal that $Na^+$-vacancy (dis)ordering in a layered oxide is intrinsically coupled to symmetry-changing phase transformations of the host structure. We examine the low-symmetry orthorhombic unit cell of P2-$Na_xNi_{1/3}Mn_{2/3}O_2$ (NNM) using both neutron and X-ray diffraction. Specifically, special sodium stoichiometries ($x = 2/3$ and $1/2$) exhibit concomitant $Na^+$-vacancy ordering and an orthorhombic distortion from the parent hexagonal unit cell. We then demonstrate that electrochemical desodiation drives symmetry-changing transformations in NNM that are linked to $Na^+$-vacancy (dis)ordering, with evidence of second-order behavior observed near $x = 2/3$. Variable-temperature synchrotron X-ray diffraction further clarifies the coupling between $Na^+$-vacancy disordering and orthorhombic-to-hexagonal phase transitions in NNM. Surprisingly, the temperature-driven phase transitions at $x = 2/3$ and 1/2 differ in character, appearing second-order and first-order, respectively. Our analysis of the phase transitions in NNM has fundamental consequences for sodium chemical diffusivity in the vicinity of the ordered phases and leads to design principles for modifying phase transition behavior in the broader class of intercalation electrodes.

## Introduction

Phase transitions that alter the symmetry of a host crystalline lattice can greatly change the properties of functional materials. A canonical example is the perovskite $BaTiO_3$ (BTO), which undergoes a symmetry-lowering cubic-to-tetragonal distortion that gives rise to ferroelectric polarization. Similarly, seminal works on $Li_7La_3Zr_2O_{12}$ (LLZO) and $Na_3Zr_2Si_2PO_{12}$ (NZSP) connect the onset of superionic conduction in these solid electrolytes to symmetry-raising phase transitions induced by heating.[1,2] In all these cases, there is an additional structural motif that is linked to the observation of phase transitions: namely, the collective displacement or ordering of an ionic species. In BTO, the $Ti^{4+}$ cations collectively displace from the centers of their octahedral coordination environments toward one of the vertices, breaking the inversion symmetry of the cubic phase.[3,4] For the superionic conductors, their respective low symmetry phases have well-characterized mobile-ion ordering ($Li^+$ and $Na^+$ for LLZO and NZSP, respectively) that is not present in the high symmetry phases.[1,2] The (dis)ordering or displacement of an ionic species on its sublattice is, thus, strongly coupled to phase transitions and emergent properties in these materials.

Phase transformations that involve mobile ionic species are important for battery intercalation electrodes,[5–8] and they are particularly prevalent among the sodium layered oxides ($Na_xMO_2$, M = transition metals).[9,10] Sodium layered oxide electrodes convert stored chemical potential energy to electrical work through simultaneous (de)insertion of $Na^+$ and faradaic charge-compensation reactions involving the $MO_2$ framework. At specific values of intercalated sodium stoichiometry ($x$), phases with highly ordered arrangements of $Na^+$ and vacancies ($V_{Na}'$ in Kröger–Vink notation) have been observed across several layered systems, as exemplified in the $Na_xMO_2$ and $Na_xMM'O_2$ families (M = Co, Mn, Ni, V and M′ = Cu, Mg).[11–19] These $Na^+$-vacancy ordered phases, and transformations between them, have been shown to impact both energy density and chemical diffusion and hence deserve careful characterization studies.[20–24]

$Na^+$-vacancy ordering is strongly coupled to the surrounding environment via electrostatic and strain interactions, among other mechanisms.[22,25,26] Thus, the $Na^+$ positions depend on both the host lattice and composition. Both transition metal composition and sodium stoichiometry, for example, significantly change observed $Na^+$ ordering in several of the $Na_xMO_2$ and $Na_xMM'O_2$ systems mentioned previously.[11,18,21,22,27–29] The $Na^+$ positions are also influenced by the stacking sequence (of either the O- or M- sublattices) of the host layered structure, as well as by stacking faults.[12,30,31] Additionally, $Na^+$-vacancy ordering is commonly observed in layered oxides with lattice distortions (symmetry lower than hexagonal).[13,32–34] However, the often simultaneous presence of $Na^+$-vacancy ordering and Jahn–Teller-active metals makes it difficult to identify their individual roles in distorting the lattice.

Recently, Pfieffer et al.[35] and Pacileo et al.[36] have reported an orthorhombic distortion at $x = 2/3$ in P2-$Na_xNi_{1/3}Mn_{2/3}O_2$ (NNM), a layered oxide that exhibits $Na^+$-vacancy ordering.[29] Interestingly, at $x = 2/3$, NNM contains no Jahn–Teller-active cations, raising the question of what drives its lattice distortion. NNM has an intricate structure, displaying compositional ordering on multiple crystallographic sublattices.[22,37,38] The transition metal layers have in-plane honeycomb ordering and predominantly stack in a –$C_1C_2$– sequence (Fig. 1a-b). In the sodiation range $1/3 < x < 2/3$, NNM adopts a P2-type oxygen stacking sequence (–ABBA–) where $Na^+$ occupy prismatic sites (P) between two-dimensional $[Ni_{1/3}Mn_{2/3}]O_2$ slabs (Fig. 1c). There are two distinct $NaO_6$ prisms that either share edges ($Na_e$) or faces ($Na_f$) with the $MO_6$ (M = $Ni^{2+}$ or $Mn^{4+}$, when $x = 2/3$) octahedra above and below it (Fig. 1c).[28] While numerous studies have examined the structure of NNM, the low-symmetry orthorhombic structure has not yet been systematically investigated.

In this work, we unambiguously show that $Na^+$-vacancy ordering in NNM is coupled to orthorhombic distortions of the host lattice. Neutron powder diffraction (NPD) and synchrotron X-ray diffraction (sXRD) confirm that as-synthesized NNM ($x = 2/3$) exhibits an orthorhombic unit cell together with substantial (~25%) stacking faults of the honeycomb-ordered transition metal layers. After identifying the orthorhombic distortion in as-synthesized NNM, we characterize its dynamic evolution. First, through electrochemical desodiation, we reveal orthorhombic-to-hexagonal phase transformations in NNM that are coupled to $Na^+$-vacancy disordering, with second-order behavior near $x = 2/3$. We then illustrate the effect of temperature on the lattice distortions: we observe a continuous orthorhombic-to-hexagonal transition at $x = 2/3$ accompanied by non-linear thermal expansion of all three lattice parameters and an underlying effective first-order transition at $x = 1/2$ in NNM. Despite distinct phase transition mechanisms, both transformations are accompanied by $Na^+$-vacancy disordering, revealing the unambiguous correlation between ion ordering and host lattice distortion. Linking $Na^+$-vacancy ordering to symmetry-changing phase transitions informs important design rules for high-rate, mechanically stable cathode materials.

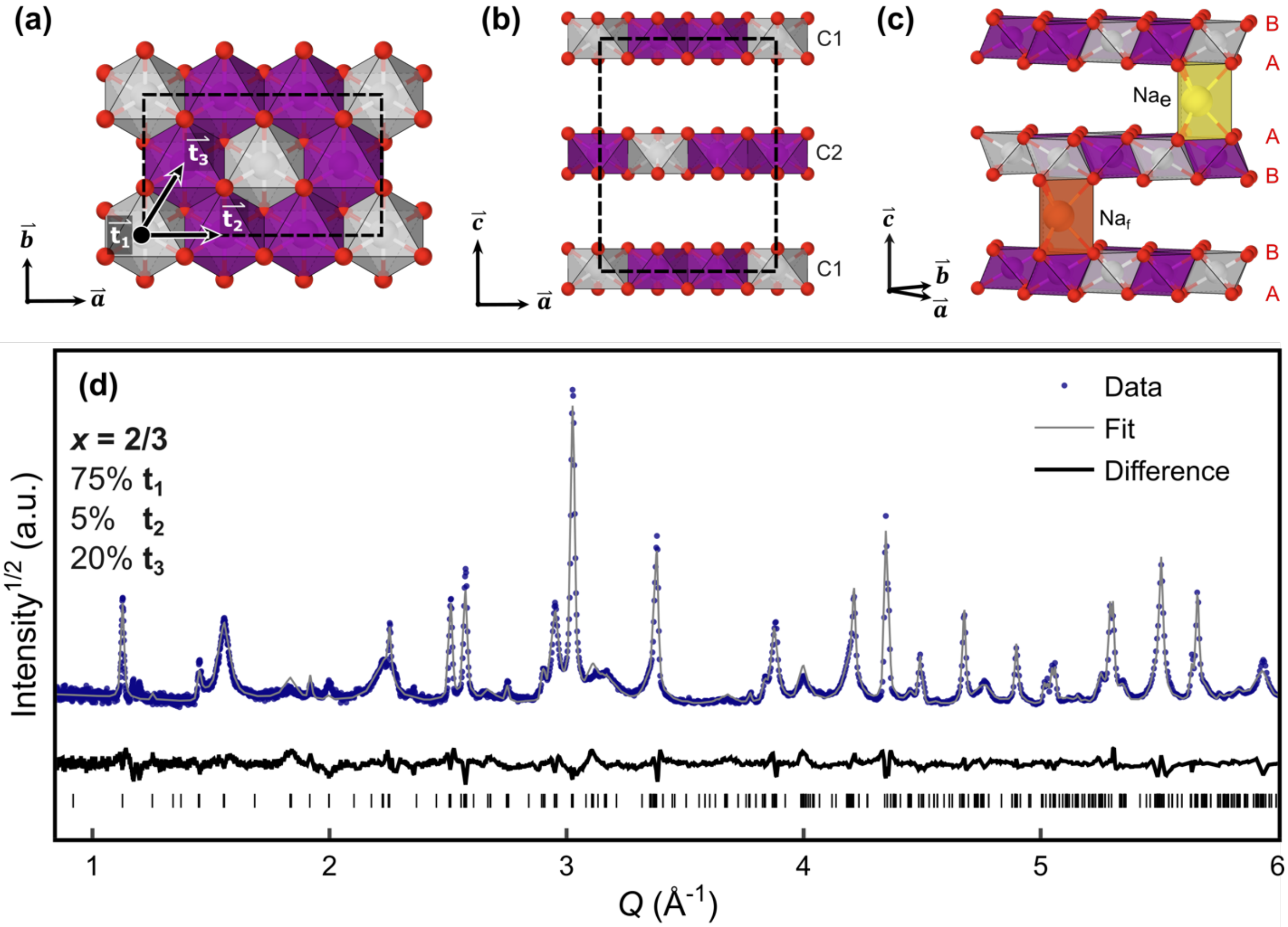

**Figure 1**. Structure of as-synthesized NNM ($x = 2/3$). (a) Ni/Mn honeycomb ordering with stacking fault vectors $\mathbf{t}_1$, $\mathbf{t}_2$, and $\mathbf{t}_3$ depicted with respect to the $P222_1$ unit cell (black dashed box). (b) Ni/Mn predominant stacking sequence (–$C_1C_2$– described by vector $\mathbf{t}_1$). (c) Depiction of the P2-type (–ABBA–) oxygen stacking sequence and the two different $Na^+$ sites ($Na_f$ and $Na_e$). Mn and Ni octahedra are shown in purple and gray, respectively, with O atoms in red. $Na_f$ and $Na_e$

atoms are orange and yellow, respectively (d) Rietveld refinement of the NPD data including stacking faults (see Figure S3 for depiction of stacking vectors). Tick marks denote calculated reflections, and $Q$ is the magnitude of the neutron scattering wavevector.

## Results

*Structural model for as-synthesized NNM*

We begin our analysis by examining the structure of as-synthesized NNM ($x$ = 2/3). The $Na^{+}$-vacancy ordering patterns for as-synthesized NNM that are calculated from first-principles methods predict several Bragg reflections that are not observed experimentally.[22,35] We note that density functional theory (DFT) predictions of ground state $Na^{+}$-vacancy orderings may be unreliable when candidate structures are close in energy because vibrational entropy is often neglected in DFT calculations and can affect relative phase stability at ambient temperature.[39] Nevertheless, here we find that transition-metal stacking faults can explain the differences between experiment and theory.

We verify the nature of the transition-metal stacking faults in NNM using NPD. These faults disrupt long-range $Na^{+}$-vacancy ordering, causing the $Na^{+}$-vacancy superlattice reflections to broaden or disappear compared to the fault-free case (compare fits in Fig. 1d and Fig. S1-S2). Starting from a model with perfect –$C_1C_2$– stacking (no faults), our refinement does not accurately capture peak profiles in the NPD pattern, particularly those with layering character ($h$ or $k \neq 0$ and $l \neq 0$) (Fig. S1 and Tables S1-S2). To build our phenomenological fault model, we incorporate the effects of $C_2C_2$ and $C_1C_2C_3$ stacking faults by including the respective stacking vectors $\mathbf{t_2}$ = [1/3, 0, 1] and $\mathbf{t_3}$ = [1/6, ½, 1]. These vectors are defined with respect to the $P222_1$ unit cell (Fig. 1a and Fig. S3). Since the repeating unit of P2-type oxides contains two distinct layers, these stacking vectors represent transitions between two-layer units (Fig. S3) rather than between each layer (which is typical for O3 or P3-type structures).[40,41] We achieve the best fit ($R_{wp}$ = 8.82%) using 5% $\mathbf{t_2}$ and 20% $\mathbf{t_3}$ (Fig. 1d and Tables S3-S4). The remaining 75% of the stacking vectors are attributed to $\mathbf{t_1}$ = [0, 0, 1], which corresponds to perfect –$C_1C_2$– stacking. We hypothesize that these stacking faults arise from partial screening by $Na^{+}$ of electrostatic interactions between adjacent transition metal layers.[38,40]

In addition to capturing the NPD peak profiles, our stacking fault model accurately describes the selective broadening of $Na^{+}$-vacancy ordering peaks with out-of-plane character (Fig.

1d). However, stacking faults have a less pronounced effect on in-plane $Na^+$-vacancy ordering peaks ($h$ or $k \neq 0$ and $l = 0$), which can be clearly seen in Figure S1. In other words, stacking faults disrupt $Na^+$-vacancy ordering between different layers in the crystal, but the in-plane ordering is preserved. These in-plane interactions impact the (de)sodiation mechanism in layered oxides and have important structural consequences that will be outlined below.

In most XRD-based studies, the structure of NNM is refined in the high symmetry hexagonal space group $P6_3/mmc$, which is sufficient for describing the P2 oxygen stacking sequence without the added complexity of stacking faults, transition metal ordering, and $Na^+$-vacancy ordering.[23] However, the choice of hexagonal symmetry is not accurate for as-synthesized NNM. Through sXRD, we observe two sets of split peaks (labeled D1 and D2 in Fig. S1) corresponding to the (100) and (110) reflections in the $P6_3/mmc$ unit cell, respectively. The splitting of these peaks is forbidden in a hexagonal setting.

Using ISODISTORT[42], we find that the simplest perturbation to the $P6_3/mmc$ unit cell that accounts for this peak splitting is an orthorhombic distortion via the irreducible representation, $\Gamma_5^+$. The orthorhombic unit cell has the space group *Cmcm*, a subgroup of the parent $P6_3/mmc$ (see Supplemental section "Refinement Details" for the transformation matrices between these two unit cells and a description of the space group selections throughout this work). This symmetry lowering renders the $a$- and $b$-lattice directions unique, leading to additional $d$-spacings that give rise to the splitting of D1 and D2 (Fig. S1). Specifically in the context of the orthorhombic unit cell in *Cmcm*, this means $b \neq \sqrt{3}a$. Throughout this work, we quantify the magnitude of distortion in the *Cmcm* unit cell ($\delta_o$) using the relationship $\delta_o = \frac{b}{\sqrt{3}a} - 1$. This parameter, $\delta_o$, is nonzero when an in-plane orthorhombic distortion is present and zero when the crystal has hexagonal symmetry. The orthorhombic distortion in as-synthesized NNM is positive ($\delta_o$ = 0.35%), indicating that $b > \sqrt{3}a$ (Fig. S4 and Table S5). Our refined orthorhombic distortion in as-synthesized NNM agrees with recent studies.[35,36]

Additionally, the orthorhombic structures need not have $Na^+$ perfectly centered on the prismatic sites, since the $\Gamma_5^+$ perturbation allows $Na^+$ to displace along the elongated axis (here, the [010] direction). Our refinement of as-synthesized NNM gives in-plane displacements of 0.21 Å and –0.14 Å for $Na_f$ and $Na_e$ sites, respectively. The proximity of $Na_f$ sites to neighboring transition metals likely drives them to displace farther than $Na_e$ sites. To the best of our knowledge, no other study has reported $Na^+$ displacements in NNM. Both the orthorhombic distortion and $Na^+$ site displacement will be important for understanding the structure of electrochemically desodiated NNM in the next section.

*Controlling symmetry in NNM via electrochemical desodiation*

Since NNM is a promising intercalation electrode, it is important to understand the evolution of the orthorhombic distortion with sodium stoichiometry. Here, we use electrochemical desodiation to control sodium stoichiometry and, simultaneously, $Na^+$-vacancy ordering in NNM. The presence or absence of $Na^+$-vacancy ordering can be inferred from the shape of the voltage profile vs. a sodium metal counter electrode.[43,44] A steep vertical region in the voltage profile typically corresponds to a single-phase material with a highly ordered crystal structure. Additionally, sloping regions are typically solid solution regimes where $Na^+$ and $V_{Na}'$ are disordered, while flat plateaus generally signify a phase transformation.

Combining electrochemical desodiation with sXRD, we characterize subtle structural changes associated with $Na^+$-vacancy (dis)ordering. We harvest five samples of NNM with different sodium stoichiometries. From Coulomb counting (assuming 100% faradaic efficiency) based on the voltage profile in Figure 2a, our sodiation states are $x$ = 2/3, 0.59, 1/2, 0.42, and 1/3 ($x$ = 2/3 is simply as-synthesized NNM). These sodium stoichiometries agree closely with those obtained from Rietveld refinements (Table S6, S7 and S9). From the voltage profile in Figure 2a, sodium stoichiometries of $x$ = 2/3, 1/2, and 1/3 are expected to have $Na^+$-vacancy ordering, while $x$ = 0.59 and 0.42 are not. In the sXRD patterns, $Na^+$-vacancy ordering peaks are best observed in the scattering vector ($Q$) range 1.4 to 2 $Å^{-1}$. This region is shown in Fig. 2b, confirming the expected presence (or absence) of $Na^+$-vacancy superlattice reflections across the harvested samples.

The $x$ = 1/2 sample can be described by a simple row-type ordering of $Na_f$ and $Na_e$ sites.[45] For this sample, unlike for $x$ = 2/3, the predicted reflections associated with the row-type ordering (space group *Pnmm*) all align with a Bragg peak measured with sXRD (Fig. S5 and Table S11), suggesting there are relatively fewer transition-metal stacking faults at $x$ = 1/2 compared to $x$ = 2/3. We hypothesize that stacking faults may be reduced at $x$ = 1/2 due to increased electrostatic repulsion between transition metal layers upon desodiation (see Fig. S6 and Supplementary

Discussion 1 for further discussion on stacking faults in this material). Notably, the orthorhombic distortion at $x$ = 1/2 ($\delta_o$ = 0.19%) is smaller in magnitude than that at $x$ = 2/3 (see peak splitting in Fig. 2c-d). Similarly, the displacements of $Na_f$ and $Na_e$ sites at $x$ = 1/2 (0.14 Å and –0.09 Å, respectively) are smaller than those refined for $x$ = 2/3 (Fig. S4 and Table S8). Thermal analysis in the following section will reveal further details about the differences between these two ordered states of NNM.

For $x$ = 1/3, it is unclear whether a distortion is present. The structural analysis of this phase (Fig. S4 and Table S10) is complicated by anisotropic peak broadening and the overlap of the D2 reflections with the (008) layering peak (arrow in Fig. 2d). Note that such features have also been observed in previous reports.[35,37] Although we obtain non-zero positive $\delta_o$ values, the sample-to-sample variability is relatively large (Table S12). Furthermore, the refinement quality is only marginally better at $x$ = 1/3 as measured by $R_{wp}$ when the data are refined in the orthorhombic space group instead of the hexagonal space group (Fig. S7). These results suggest that a distortion is negligible for $x$ = 1/3 (see Supplementary Discussion 2 for further details). Another consideration is that the $Na^+$-vacancy ordering configuration at $x$ = 1/3 maintains in-plane hexagonal symmetry,[29] while our models at $x$ = 2/3 and 1/2 are orthorhombic (see Fig. 2e and Fig. S8). Given the direct correlation between the symmetry of the mobile ion sublattice and that of the host structure, we hypothesize that $Na^+$-vacancy ordering can itself drive the host lattice distortion. We note that the $Na^+$ displacement from the centers of the high-symmetry prismatic sites at $x$ = 2/3 and 1/2 increases the average distance between neighboring $Na^+$ ions (see, for example, Fig. S5), which can be attributed to repulsive $Na^+$-$Na^+$ forces. In contrast, the hexagonal $Na^+$-vacancy ordering of $x$ = 1/3 already minimizes repulsive interactions between $Na^+$.

Importantly, NNM samples harvested in disordered states ($x$ = 0.59 and 0.42) exhibit a clear increase in symmetry. The sXRD patterns do not show any orthorhombic peak splitting (Fig. 2c-d), suggesting that the long-range orthorhombic distortion has disappeared ($\delta_o$ = 0) in these samples and they have average hexagonal symmetry. The presence of measurable distortions in samples with $Na^+$-vacancy ordering that breaks in-plane hexagonal symmetry and their absence otherwise reveals a fundamental coupling between $Na^+$-vacancy ordering and host lattice symmetry. These results are summarized schematically in Figure 2e.

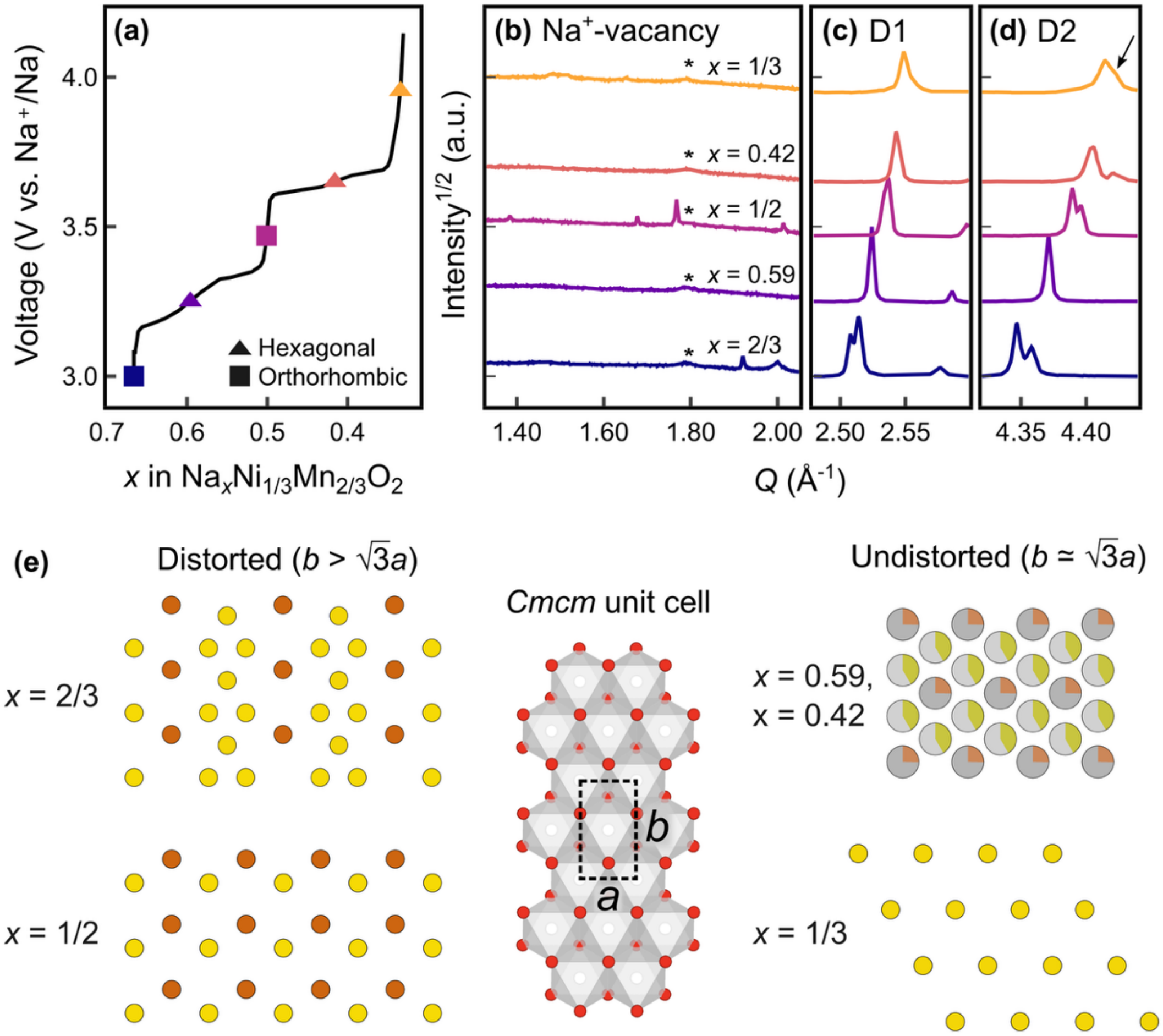

**Figure 2.** *Ex situ* sXRD of electrochemically desodiated NNM. (a) Voltage profile of NNM from $x = 2/3$ to $x = 1/3$ measured at a constant current of 17.3 mA $g^{-1}$, with markers showing the cut-off voltages at which samples were harvested. Squares indicate an orthorhombic unit cell, while triangles represent samples with hexagonal symmetry ($\delta_o = 0$). (b)–(d) *Ex situ* sXRD highlighting $Na^+$-vacancy ordering, D1, and D2 reflections, respectively. (e) Schematic summarizing sXRD results, with $Na_f$, $Na_e$, O, and transition metals depicted in orange, yellow, red, and gray, respectively (noting we cannot distinguish Ni and Mn with the *Cmcm* model). Intensity scaling in (b) is 2.5x that in (c)–(d) to emphasize the subtle $Na^+$-vacancy ordering peaks. Asterisks in (b) indicate an unknown impurity present in all samples. Arrow in (d) indicates the NNM (008) peak. In (e), $Na^+$-vacancy ordering model at $x = 0.59$ and 0.42 is illustrative, while the model at $x = 1/3$ was reported in previous literature.[29] $Q$ is the magnitude of the X-ray scattering wavevector.

Next, we monitor the D2 peaks using *operando* XRD on a lab diffractometer to characterize the dynamic crossover between orthorhombic and hexagonal symmetry with desodiation. Distinct regimes of crystallographic D2 peak splitting can be seen clearly in Figure 3a, despite the use of the lab source resulting in Cu K$\alpha_1$ and K$\alpha_2$ doublets for individual Bragg peaks. Since the intensity ratio of Cu K$\alpha_1$ and K$\alpha_2$ radiation is fixed at 2:1, any deviation in the intensity ratio indicates peak splitting due to the orthorhombic structure. For additional clarity, Figure S9 shows the individual *operando* XRD scans corresponding to the sodiation states of the harvested samples presented in Figure 2. The crossover of the regimes in D2 correlate with voltage plateaus in Figure 3a, signifying orthorhombic-to-hexagonal phase transformations that are driven by changing sodium stoichiometry. Over $0.625 < x < 2/3$, we refine a continuous change in the orthorhombic distortion from ~0.3% to zero, consistent with a second-order phase transition mechanism (see Fig. 3a, Fig. S9, and Supplemental section "Refinement Details"). While the transformation near $x = 1/2$ is also apparent beginning from $x \sim 0.54$, we were unable to refine the orthorhombic distortion with this dataset.

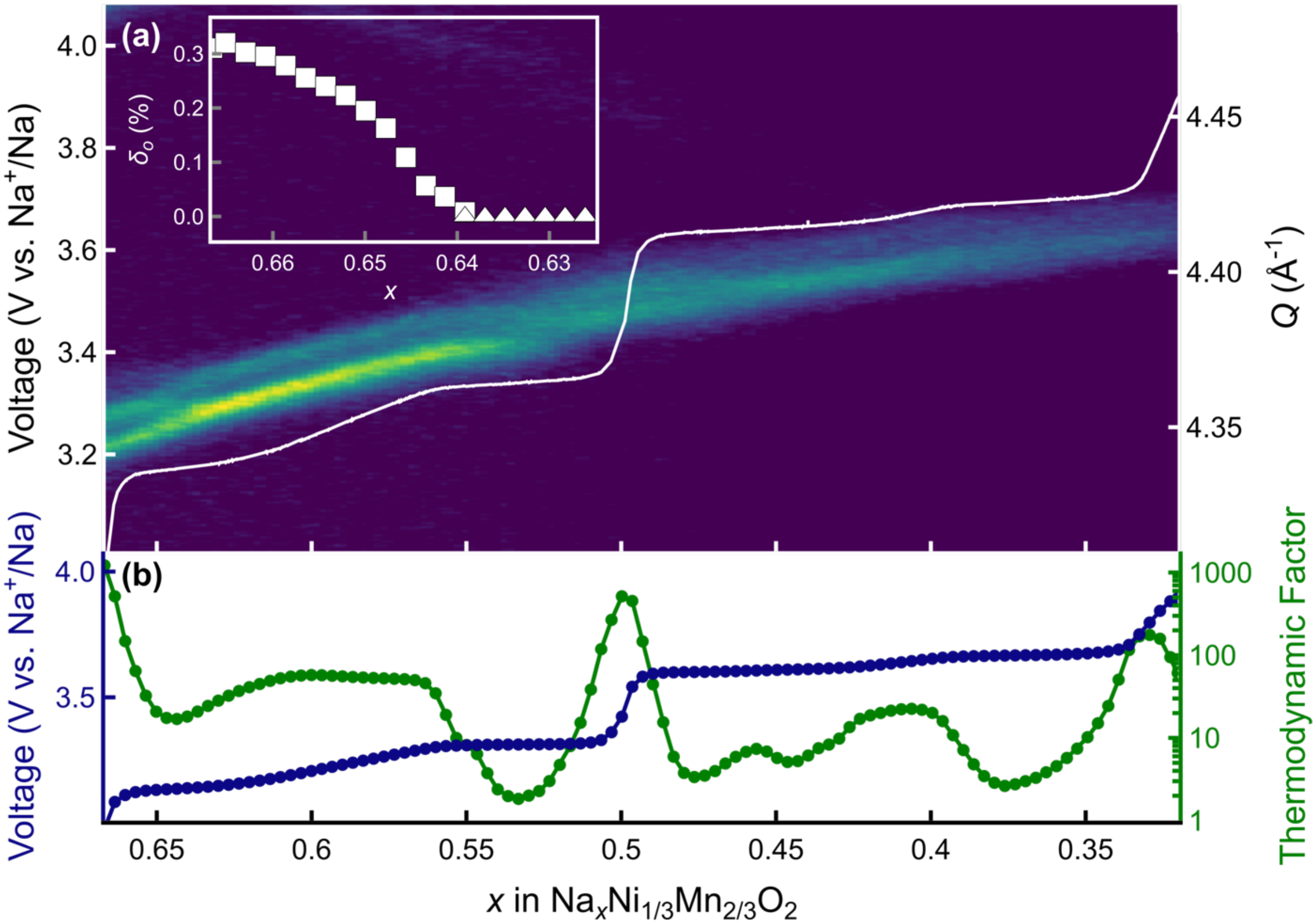

**Figure 3.** *Operando* XRD monitoring orthorhombic-to-hexagonal phase transformations in NNM via the D2 peaks. (a) *Operando* XRD heatmap on a lab source (with both Cu K$\alpha_1$ and K$\alpha_2$ radiation) tracking the orthorhombic distortion with simultaneous $Na^+$-vacancy ordering monitored via the voltage profile measured at a constant current of 3.5 mA $g^{-1}$ (white curve). $Q$ is the magnitude of the X-ray scattering wavevector. Inset in (a) shows the continuous change in the orthorhombic distortion over the sodium stoichiometry range $0.625 < x < 2/3$, with squares and triangles representing orthorhombic and hexagonal structural symmetry, respectively. (b) Pseudo open circuit voltage profile (blue trace) plotted alongside the thermodynamic factor (green trace). Raw intensity (rather than square root intensity) scaling is used for the *operando* XRD heatmap in (a).

The electrochemically-induced phase transformations in NNM have direct implications for the sodium chemical diffusion coefficient, which can be inferred from the voltage profile. Specifically, this correlation can be quantified through the thermodynamic factor ($\theta$), given by[44]

$$\theta = -\frac{F}{RT}\left(\frac{\partial V_{oc}}{\partial \ln(x)}\right).$$

Here, $F$ is Faraday's constant, $R$ is the ideal gas constant, $T$ = 303 K, and $V_{oc}$ is the (pseudo) open circuit voltage at $x$ (blue trace in Fig. 3b, see Supplemental section "Methods" for more details). The thermodynamic factor links the chemical diffusion coefficient to gradients in chemical potential instead of concentration, since it is the chemical potential gradient that drives diffusion.[46,47] The thermodynamic factor becomes especially important in nondilute solid-solutions and during second-order phase transitions, where the chemical potential gradient deviates from the concentration gradient. Figure 3b shows that $\theta$ changes continuously by almost two orders of magnitude over the range $0.625 < x < 2/3$, suggesting a link between this phase transformation and the sodium chemical diffusion coefficient in NNM.

Approaching the critical point of a second-order transition (here, $x \sim 0.625$), the coherence length of fluctuations in the $Na^+$ density diverges, and it becomes progressively harder to drive net $Na^+$ flux.[47] This phenomenon of critical slowing down may result in a significantly reduced chemical diffusion coefficient in the vicinity of the order-disorder phase transitions in NNM (i.e., at the local minima in the thermodynamic factor in Fig. 3b). We emphasize that this is distinct from the ion mobility, which is independent of the thermodynamic factor. An important caveat to this discussion is that the thermodynamic factor is discontinuous during first-order transitions, which will become relevant when we examine the temperature-induced transitions in NNM, particularly for $x$ = 1/2, in the next section.

*Temperature-induced orthorhombic-to-hexagonal transitions in NNM*

We continue our investigation of the coupled $Na^+$-vacancy ordering and symmetry-changing phase transitions in NNM using variable temperature sXRD. Figure 4 shows the temperature-dependent sXRD on heating and cooling for the $x$ = 2/3 sample, with lattice parameters extracted via sequential Rietveld refinement in the *Cmcm* space group (see Figs. S10-S12 and Tables S13-S14 for sample refinements). The loss of the $Na^+$-vacancy ordering peaks is closely correlated with an orthorhombic-to-hexagonal phase transition with a critical temperature ($T_c$) of 310 ºC. This observed critical temperature aligns with previous observations of the temperature-dependent loss of $Na^+$-vacancy ordering in as-synthesized NNM.[38,48]

As with the composition-induced phase transition, the thermally-induced phase transition is also continuous with temperature at $x$ = 2/3. The *a*- and *c*-parameters expand monotonically but nonlinearly throughout the transition, especially the $a$-parameter (Fig. 4d). Notably, the $b$-parameter exhibits negative thermal expansion, contracting by ~0.05% between 250 and 310 ºC (Fig. 4e). Consistently, we refine a decrease of $\delta_o$ from ~0.3% to zero (Fig. 4g). In calorimetry, the phase transition exhibits a broad peak in specific heat capacity between approximately 250 ºC and 340 ºC (Fig. S13). The continuous nature of the phase transition, lack of observable hysteresis on heating and cooling, and calorimetry results provide evidence that it is second-order. Plotting the order parameter (here, $\delta_o$) against temperature on a log-log scale further demonstrates the second-order behavior near the critical temperature of 310 ºC (Fig. S14).

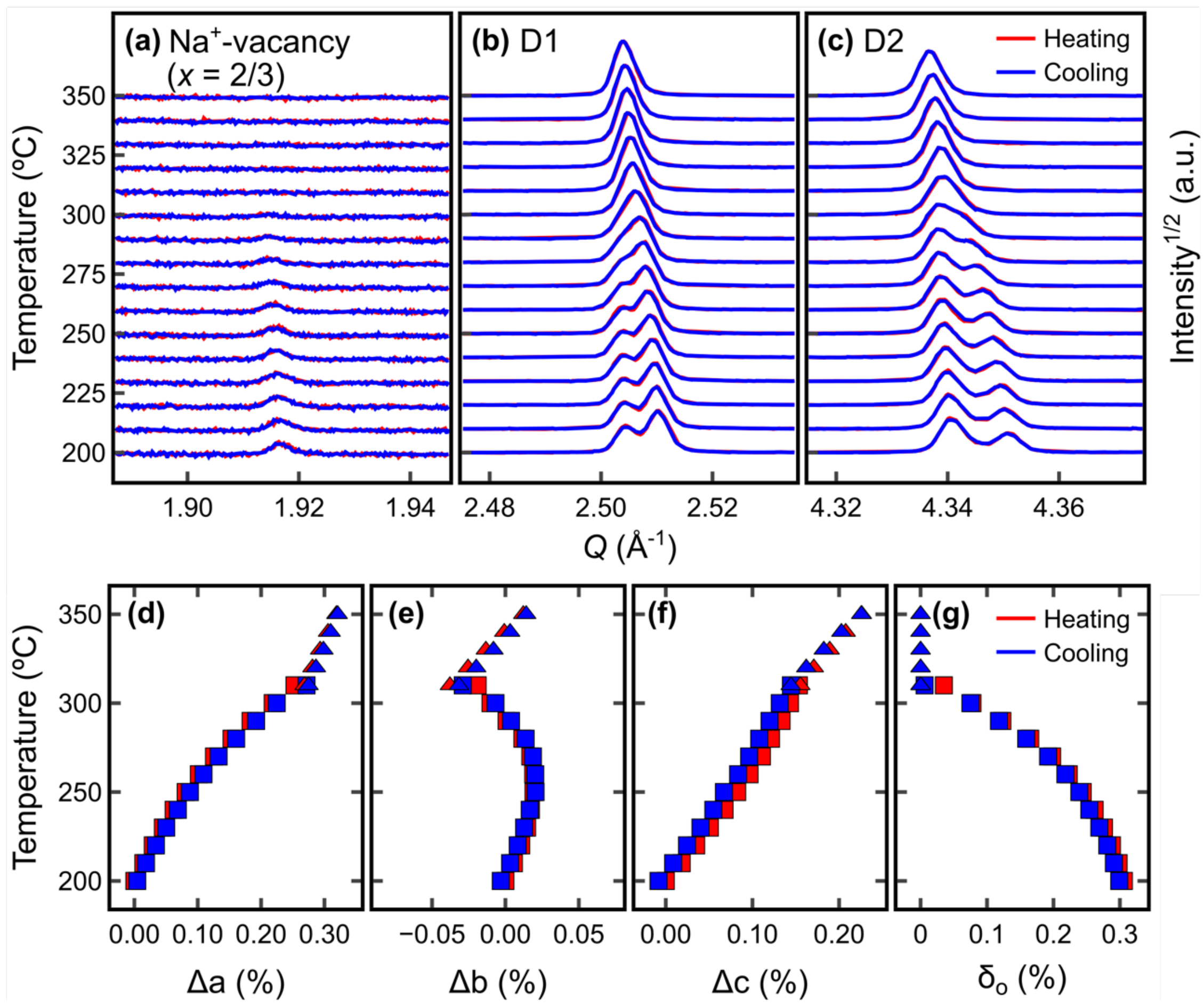

**Figure 4.** *In situ* variable temperature sXRD of as-synthesized NNM ($x$ = 2/3) on both heating (red traces) and cooling (blue) between 200 ºC and 350 ºC. (a)–(c) $Na^+$-vacancy ordering region of the sXRD pattern, D1, and D2 peaks, respectively. $Q$ is the magnitude of the X-ray scattering wavevector. (d)–(f) Temperature-dependent percentage change (Δ) of $a$-, $b$-, and $c$-parameters of the *Cmcm* unit cell with respect to the first measured values on heating, respectively. (g) Evolution of the orthorhombic distortion parameter ($\delta_o$) with temperature. Intensity scaling in (a) is 4.5x that in (b)–(c) to emphasize the $Na^+$-vacancy ordering peaks. Error bars in (d)–(g) are smaller than the size of the markers. Squares indicate orthorhombic symmetry was used in the refinement, while triangles indicate hexagonal symmetry was used. A 10 ºC $min^{-1}$ ramp was used during heating and cooling, with a one-minute equilibration prior to starting a measurement.

At $x = 1/2$, we also observe $Na^+$-vacancy disordering together with an orthorhombic-to-hexagonal phase transition (Fig. 5a-c). If we refine the data in Figure 5 assuming a continuous transition, as was the case for $x = 2/3$, we find a lower critical temperature of 175 ºC (Fig. S15). However, the continuous phase transition assumption is not valid at $x = 1/2$. Instead, we observe subtle peak splitting in the (00*l*) peaks that reveals an underlying first-order mechanism. This is most clearly observed in the higher-order (00*l*) peaks (compare Fig. 6a and 6b), which show a two-phase coexistence between 105 ºC and 175 ºC. The emergence of the high-temperature phase coincides with an intermittent decrease in intensity of the D1 peak as the volume fraction of the orthorhombically distorted phase decreases (Figure 6c). In contrast, the $x = 2/3$ phase does not show any such two-phase coexistence across the phase transition range (Fig. 6d-f). Due to the small difference in lattice parameters between the low- and high-temperature phases, a two-phase Rietveld refinement could not be carried out at $x = 1/2$.

Additionally, the $x = 1/2$ sample degraded above ~190 ºC, likely due to reaction with binder (polyvinylidene fluoride) and conductive carbon. This degradation is evidenced by anomalous contraction of the *c*-parameter above 190 ºC (Fig. S15c) and a decrease in area of the (002) layering peak (Fig. S16). We highlight that this degradation begins after the phase transformation has completed (~175 ºC), indicating that the observed phase transition is intrinsic to NNM (see Supplemental Methods and Fig. S16 for more details). As a result, high quality cooling data could not be measured to test the reversibility of this phase transition.

The fact that the temperature-induced phase transitions at $x = 2/3$ and 1/2 proceed via different mechanisms could lead to design principles for modifying phase transition behavior in cathodes. Continuous or second-order intercalation reactions are desirable since they can occur with minimal hysteresis and avoid the mechanical strain at phase boundaries associated with first-order transitions.[9] We highlight that the $x = 2/3$ and 1/2 phases differ in (1) ionic structure ($Na^+$-vacancy ordering), (2) electronic structure (Jahn–Teller-active cations), and (3) stacking fault content. Each of these can influence the phase transition pathway, such as through electron-lattice coupling in the case of the Jahn–Teller effect.[49] Stacking faults and other defects can also broaden an intrinsically first-order transition by creating a distribution of local transition temperatures (or voltages).[50,51] We emphasize that stacking fault content can be controlled through synthesis, which could make it a useful mechanism to modify phase transitions.[40,41,52] Future work combining experiment and theory is needed to clarify the underlying reasons for the observed discrepancies between the $x$ = 2/3 and 1/2 phases of NNM.

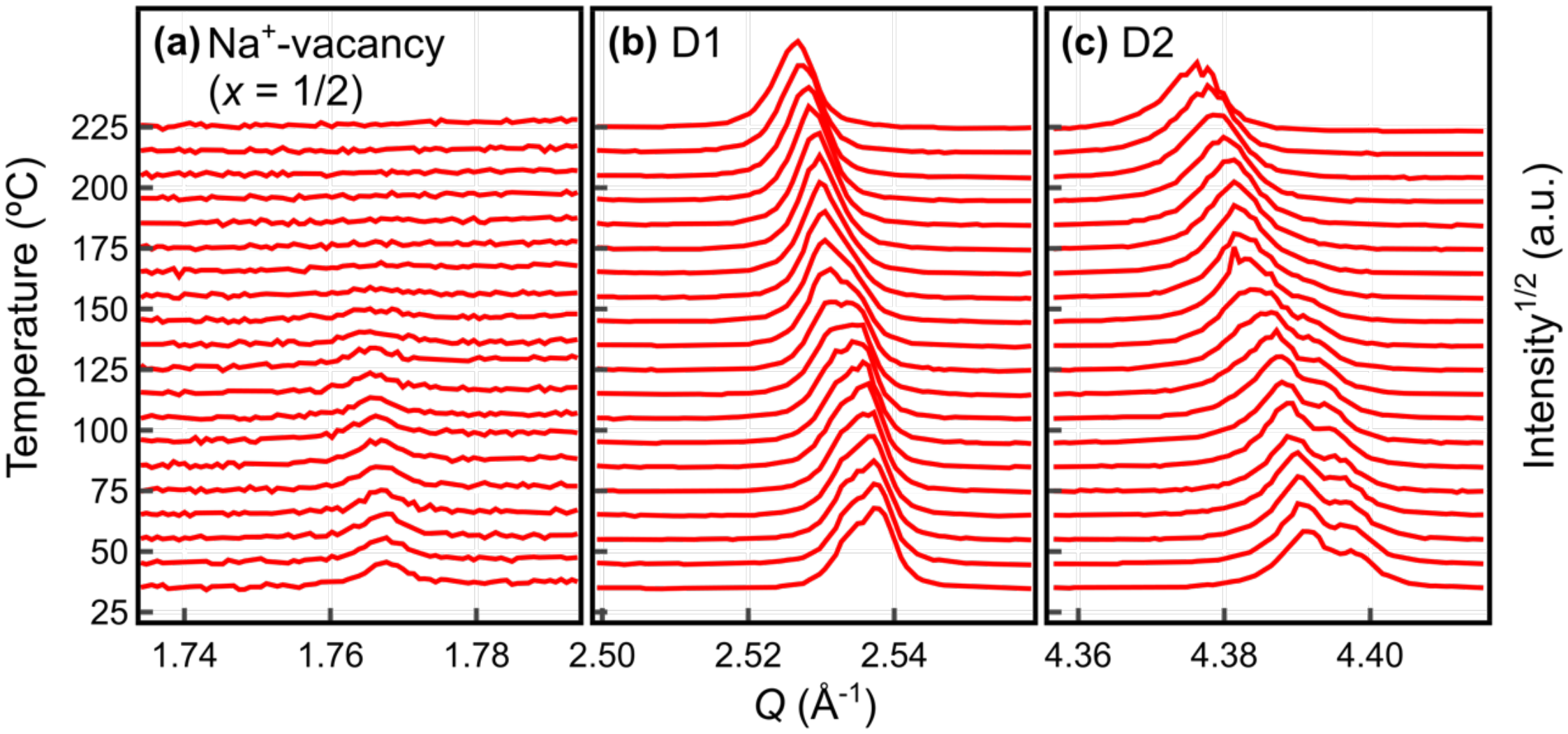

**Figure 5.** *In situ* variable temperature sXRD of the $x$ = 1/2 phase on heating from 35 ºC to 225 ºC. (a)–(c) $Na^+$-vacancy ordering region of the sXRD pattern, D1, and D2 peaks, respectively. $Q$ is the magnitude of the X-ray scattering wavevector. Intensity scaling in (a) is 3.1x that in (b)–(c) to emphasize the $Na^+$-vacancy ordering peaks. A 10 ºC $min^{-1}$ ramp was used during heating, with a one-minute equilibration prior to starting a measurement.

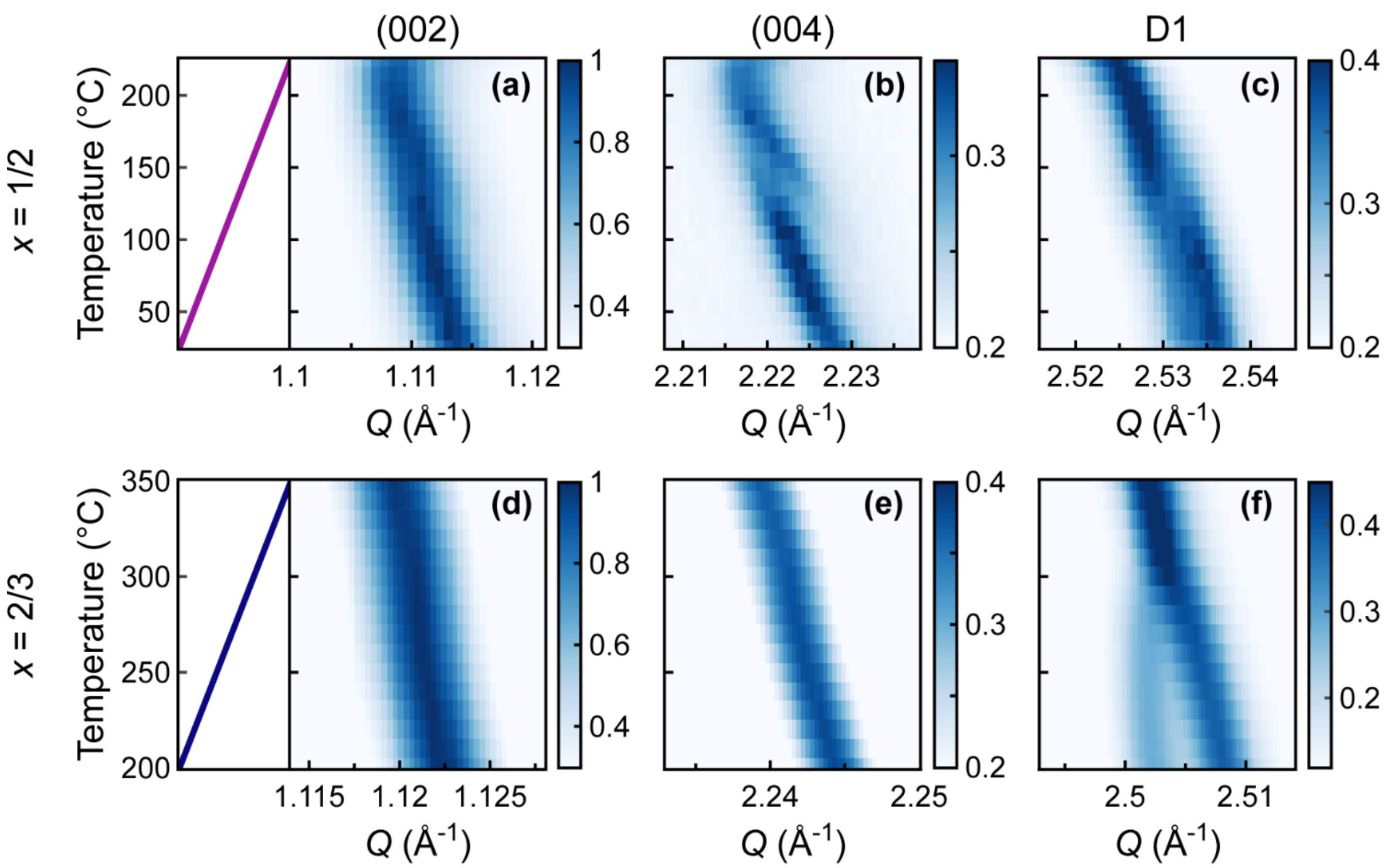

**Figure 6.** Variable temperature sXRD phase transition comparison. (a)-(c) The (002), (004), and the D1 peak, respectively, for the *x* = 1/2 electrochemically desodiated phase showing two-phase coexistence. (d)-(f) The (002), (004), and the D1 peak, respectively, for the *x* = 2/3 phase showing a continuous phase transition. Intensities are normalized to the maximum intensity of the respective (002) peak (a, d). Temperature ramps are shown on the left.

## Discussion

In this work, we have identified an intrinsic link between $Na^+$-vacancy disordering and orthorhombic-to-hexagonal phase transitions in NNM. The low symmetry structures have several distinguishing features, including (1) long-range $Na^+$-vacancy ordering, (2) an in-plane orthorhombic distortion, and (3) collective displacement of $Na^+$ away from the high symmetry $Na_f$ and $Na_e$ sites. The high symmetry phase has average hexagonal symmetry and occurs when $Na^+$ and $V_{Na}'$ are disordered. We have also demonstrated how the orthorhombic-to-hexagonal phase transitions in NNM are induced by both electrochemical desodiation and heating.

We speculate that these symmetry-changing phase transitions in NNM are driven by the sodium sublattice itself. We show in Figures 2e and S8 that only $Na^+$-vacancy ordering configurations that cannot be described by in-plane hexagonal unit cells had measurable distortions (greater than sample-to-sample variation). This suggests that the symmetry of the sodium sublattice determines the symmetry of the host lattice. When the sodium sublattice is ordered, the structure is rigid, with $Na^+$ locked in asymmetric potential wells. Then, either by changing the $V'_{Na}$ stoichiometry or adding thermal energy, the sodium sublattice relaxes to average hexagonal symmetry, and so too does the host structure. With these insights, the symmetry of a mobile sublattice could serve as a powerful driver of phase transitions in a broader class of materials beyond NNM. However, we first note that our $Na^+$-vacancy model for $x = 2/3$ is slightly different than what has been proposed in previous computational reports in which a hexagonal model is used.[23,35] Additionally, regarding the $x = 1/3$ phase, it is possible that hexagonal symmetry could be broken along the stacking direction, resulting in measurable distortions that emerge below room temperature. Like with the $x = 2/3$ phase, stacking faults at $x = 1/3$ preclude accurate refinement of symmetry elements along the stacking direction.[37] It will be important to clarify the true ground state orderings at $x = 2/3$ and 1/3 in future computational work.

Future work is also needed to better understand the phase transition mechanisms in NNM. While we report second-order-like behavior at $x = 2/3$, it is possible for a first-order transition to involve a continuous change in symmetry. For example, a two-phase, three-component system could evolve continuously in composition along a tie line such that the structural symmetry appears to change continuously with sodium stoichiometry, even though the material is traversing a two-phase region. This is unlikely at $x = 2/3$ in NNM, however, as it would require a third component (beyond Na and $MO_2$) to be sufficiently mobile. Additionally, intraparticle strain at a phase boundary or non-equilibrium charging rates could make a first-order transition appear continuous.[53] On the experimental side, future work should explore the effects of stacking faults on modifying phase transition behavior in NNM and other intercalation electrodes that display similar phase transition behavior.

Finally, while this study captures the average structures of NNM via time- and bulk-averaged diffraction measurements, a description of the local picture is needed to elucidate the phase transition mechanisms and ultimately understand their relationships to diffusivity. In particular, the $Na^+$ in hexagonal NNM may be statically disordered, harmonically/anharmonically displacing about high symmetry sites, or dynamically hopping between symmetry-related displaced sites. That is, while the average structure is hexagonal, short-range low-symmetry $Na^+$ arrangements may persist past the critical points at $x = 2/3$ and 1/2. A simulated annealing study of similar layered systems with Jahn–Teller distortions points to dynamic hopping between symmetry-related sites as being particularly likely,[34] and direct evidence of this mechanism has recently been observed in BTO.[3] Further simulated annealing and structural relaxation computations on NNM at various values of $x$ would help to clarify the local $Na^+$ environments as the phase transition proceeds between the low- and high-symmetry phases. Such work would also elucidate changes to the coherency and spatial distribution of $Na^+$-vacancy ordered crystallites as the critical transition temperatures are approached.

**Supporting Information**

Material synthesis, electrochemical desodiation methods, neutron and X-ray powder diffraction methods and data analysis; operando X-ray electrochemistry methods; methods and results for ICP-OES, differential scanning calorimetry, FTIR, and SEM; space group relationships; Rietveld refinement tables and figures; stacking fault modeling and further discussion on how stacking faults influence XRD and NPD peak shapes; further discussion on the absence of orthorhombic distortion in ordered $Na_{1/3}Ni_{1/3}Mn_{2/3}O_2$.

**Acknowledgements**

This work was supported by the Office of Naval Research under contract number N00014-23-1-2310. D.A.E. was partially supported by a National Science Foundation Graduate Research Fellowship under grant number 2146755. K.J.H. acknowledges support from the US Department of Energy, Office of Basic Energy Sciences, Division of Materials Sciences and Engineering under contract no. DE-AC02-76SF00515. J.C. gratefully acknowledges the Lindemann Trust for funding. Use of the Stanford Synchrotron Radiation Lightsource (SSRL), SLAC National Accelerator Laboratory, is supported by the U.S. Department of Energy, Office of Science, Office of Basic Energy Sciences under Contract No. DE-AC02-76SF00515. A portion of this research used resources at the Spallation Neutron Source, a DOE Office of Science User Facility operated by the Oak Ridge National Laboratory. The beam time was allocated to POWGEN on proposal number IPTS- 33714.1. Part of this work was performed at nano@stanford RRID:SCR_026695. We acknowledge K. H. Stone and M. B. Preefer for experimental support at SSRL BL 2–1. We thank Prof. P. G. Khalifah for valuable discussions on interpretation of sXRD data. Additionally, we thank K. Frohna and L. Stolz for their contribution to the design of the *operando* coin cell. We are also grateful for valuable discussions with G. M. Busse, E. P. Choy, D. Eum, E. Kaeli, N. Kapate, N. Liang, A. Lindeman, S. Patil, and D. F. Rivera about our results.

**Competing Interests**

The authors declare no competing interests.

## Data Availability

All powder diffraction from this work can be accessed at our Zenodo repository (https://doi.org/10.5281/zenodo.19135423).

**Supporting information for "Electrochemical and thermal control of continuous phase transitions in P2-$Na_xNi_{1/3}Mn_{2/3}O_2$"**

Dylan A. Edelman[1], John Cattermull[1], Jue Liu[2], Zhelong Jiang[1,3], Hari Ramachandran[1], Edward Mu[4], Cheng Li[2], Anton Van der Ven[5], Katherine J. Harmon[1*], William C. Chueh[1,6,7*]

*[1]Department of Materials Science and Engineering, Stanford University, Stanford, CA, 94305, USA.*

*[2]Neutron Scattering Division, Oak Ridge National Laboratory, Oak Ridge, TN, 37831, USA.*

*[3]SLAC-Stanford Battery Center, Applied Energy Division, SLAC National Accelerator Laboratory, Menlo Park, CA, 94025, USA.*

*[4]Department of Chemistry, Stanford University, Stanford, CA, 94305, USA.*

*[5]Materials Department, University of California, Santa Barbara, Santa Barbara, CA, 93117, USA.*

*[6]Stanford Institute for Materials and Energy Sciences, SLAC National Accelerator Laboratory, Menlo Park, CA, 94025, USA.*

*[7]Department of Energy Science and Engineering, Stanford University, Stanford, CA, 94305, USA.*

**Corresponding author e-mails: kharmon@stanford.edu, wchueh@stanford.edu*

## Methods

### *i. Synthesis and electrochemical characterization*

$Na_{2/3}Ni_{1/3}Mn_{2/3}O_2$ was synthesized by a solid-state reaction. Stoichiometric amounts of NiO (99 wt. % trace metals basis, Alfa Aesar) and $MnCO_3$ (99.9 wt. %, Sigma-Aldrich) were first ball-milled in ethanol with 6 mol.% excess $Na_2CO_3$ (99.5 wt. %, Sigma-Aldrich) at 300 rpm for 3 hours in intervals of three minutes followed by a one-minute rest, with direction reversal every interval. The mixture was then transferred to a hotplate to dry overnight. The dried material was subsequently ground with a mortar and pestle, pelletized (25 mm diameter), and calcined in air in a box furnace at 950 ºC for 14 hours with a heating ramp rate of 5 ºC $min^{-1}$. The resulting powder was furnace-cooled to room temperature before it was transferred to an argon-filled glovebox. The metal stoichiometry of the final product was measured with inductively coupled plasma optical emission spectroscopy (ICP-OES) to be $Na_{0.70}Ni_{0.34}Mn_{0.66}O_2$ (assuming a stoichiometry of $Na_xNi_yMn_{1-y}O_2$), closely matching the target stoichiometry. The excess Na is likely not incorporated into the NNM structure and rather exists as residuals on the surface, such as carbonate species[1] as evidenced by carbonated peaks observed via Fourier transform infrared spectroscopy (Fig. S17). The morphology of NNM, imaged via scanning electron microscopy (SEM), is shown in Figure S18.

To obtain the electrochemically desodiated phases for *ex situ* synchrotron X-ray diffraction (sXRD), we first prepared slurries of 80 wt.% $Na_{2/3}Ni_{1/3}Mn_{2/3}O_2$ active material, 10 wt.% conductive carbon (Super-P, Alfa Aesar, H30253), and 10 wt.% polyvinylidene fluoride binder (PVDF, Kynar HSV900) dissolved in N-methyl-2-pyrrolidone (Sigma-Aldrich, >99.5% anhydrous). The slurries were cast onto carbon-coated aluminum foil with a doctor blade at a height of 200 μm prior to being dried overnight in a vacuum oven at 60 ºC. Typical active material loadings were ~5 mg $cm^{-2}$. Next, the electrodes were punched into 1 $cm^2$ discs and assembled in coin cells (CR2032, Wellcos) using a glass fiber separator (GF/F, Whatman) and a sodium metal counter electrode. We used approximately 80 μL of electrolyte (1M $NaPF_6$ in ethylene carbonate:diethyl carbonate:dimethyl carbonate in a 1:1:1 volume ratio with 1% fluoroethylene carbonate, MTI) in every cell.

Galvanostatic cycling for the harvested samples was performed at 30 ºC using a Biologic BCS-805 battery cycler by applying a current of 17.3 mA $g^{-1}$ until the cell reached the desired cut-off voltage, at which point a five-hour voltage hold was applied to ensure the electrode reached a uniform sodiation state. The cells were then immediately transferred to an argon-filled glovebox and disassembled. The electrodes were subsequently washed in excess dimethyl carbonate (DMC) solvent, scraped, and packed into capillaries. Glass capillaries (0.5 mm outer diameter and 0.01 mm wall thickness, Charles Supper Co.) were used for room temperature sXRD measurements, while quartz capillaries of the same dimensions were used for variable temperature sXRD. The glass capillaries were sealed with vacuum grease inside an argon-filled glovebox and then subsequently flame sealed outside the glove box, while the quartz capillaries were sealed inside the glovebox with epoxy (Loctite Instant Mix 5 Minute).

To obtain the (pseudo) open circuit voltage profile for calculation of the thermodynamic factor in main text Figure 3b, we performed the galvanostatic intermittent titration technique. We applied current pulses of 17.3 mA $g^{-1}$ for approximately three minutes, followed by a one-hour rest. We then took the open circuit voltage to be the average of the last 20 points (10 minutes) recorded during the rest.

*ii. Neutron powder diffraction (NPD) measurement*

Approximately 0.8 g of $Na_{2/3}Ni_{1/3}Mn_{2/3}O_2$ powder was loaded into a 6 mm diameter vanadium can and sealed inside an argon-filled glovebox prior to measurement at the POWGEN diffractometer (BL-11a) at the Spallation Neutron Source at Oak Ridge National Laboratory. Measurement was performed using the 1.5 Å constant-wavelength frame at room temperature. Instrumental parameters were determined by fitting to the measured diffraction patterns of a NIST Si 640d standard reference material.

*iii. Synchrotron x-ray diffraction (sXRD) and lab-scale XRD measurement*

*Ex situ* capillary transmission sXRD was performed at beamline 2-1 at the Stanford Synchrotron Radiation Lightsource (SSRL) using a wavelength of 0.7296305(6) Å at ambient temperature. The scan range was 3 to 83.8° ($2\theta$) with step size 0.005°, and each scan was approximately four minutes in duration. The capillaries with sample were continuously rotated during data collection to avoid preferential orientation effects. Raw data were recorded using a Pilatus100K detector at a distance of 700 mm from the sample. High-throughput capillary sXRD measurements were performed using the robotic sample exchanger.[2]

The variable temperature capillary transmission sXRD measurements were also performed at beamline 2-1 at SSRL using an Anton Paar HTK 1200N temperature chamber. The wavelength was the same as the *ex situ* measurements. The scan range was 3 to 63.8° ($2\theta$) with step size 0.005°, and each scan was approximately three-and-a-half minutes in duration. Quartz capillaries were used inside the temperature chamber to avoid warping at elevated temperatures. A ramp rate of 10 °C $min^{-1}$ was applied between scans. Before starting a new measurement, the temperature was held at the set point for at least one minute for equilibration.

*Operando* XRD measurements were conducted at ambient temperature in a 2032-type coin cell with a Kapton window sealing an 8 mm diameter laser-cut hole in the cathode-side stainless steel casing. Steel mesh was used as the current collector in place of carbon-coated aluminum. A low current magnitude of 3.5 mA $g^{-1}$ was applied using a Biologic SP-200 potentiostat to measure near-equilibrium behavior. The measurements were performed in reflection mode using an Empyrean diffractometer equipped with a Cu source, yielding Cu K$\alpha_1$ and Cu K$\alpha_2$ radiation. The scan range was 63.5 to 67.0° ($2\theta$) with step size 0.0072°, with a small range chosen to maximize the time resolution between measurements. Individual scans were approximately 10 minutes in duration. Subsequent scans were added together in post-processing to improve the signal-to-noise ratio.

### *iv. Inductively coupled plasma optical emission spectroscopy (ICP-OES)*

ICP-OES was used for composition determination. ICP-OES was performed by EAG Laboratories in Liverpool, New York using a Perkin Elmer Optima 7300V instrument. The as-synthesized NNM sample was weighed and digested using both hydrochloric and nitric acid. An addition of 0.05 mL of a 1000 mg/L Sc solution was used as internal standard before being brought to 50 mL final volume with deionized water.

A blank and three standard calibration solutions were prepared bracketing the samples. The samples were evaluated against a linear curve, having a minimum calibration coefficient of 0.999. A second source quality control solution was prepared at a concentration falling within the range of the standard curve for all elements. This quality control standard was run at the beginning and end with an acceptance criterion of ±10%.

The sample was converted to an aerosol through a nebulizer (Teflon GemCone) and directed to an argon plasma, generating emission spectra. The characteristic wavelengths of the resulting light emissions were then measured to determine the mass ratio of Na, Ni, and Mn in the NNM sample. These values were converted to a stoichiometry of $Na_{0.70}Ni_{0.34}Mn_{0.66}O_2$, assuming a stoichiometry of $Na_xNi_yMn_{1-y}O_2$.

### *v. Differential scanning calorimetry*

Differential scanning calorimetry (DSC) was performed in a DSC Q2500 (TA Instruments). Approximately 5 mg of NNM powder was loaded into a Tzero aluminum pan and spread in a thin layer prior to sealing with the lid. A blank Tzero aluminum pan was used as a reference. The heating rate for the DSC measurement was 10 ºC $min^{-1}$. Data analysis was performed directly in TRIOS software to compute heat capacity as a function of temperature.

### *vi. Fourier transform infrared spectroscopy*

Fourier transform infrared spectroscopy (FTIR) was performed in a Thermo Scientific Nicolet 6700 with an attenuated total reflectance attachment. A background scan was collected in air prior to measuring under the same conditions with NNM powder.

### *vii. Scanning electron microscopy*

Scanning electron microscopy was performed in a FEI Magellan 400 XHR equipped with a Field Emission Gun. NNM powder was dispersed in ethanol before drop casting onto a stainless-steel sample mount. An accelerating field of 2 kV and probe current of 25 pA were used for imaging.

## Refinement Details

All structural refinements in this work were performed using TOPAS software.[3] The refinement strategy differed between samples, and the details of all refinements are outlined below. The wavelength and instrumental parameters were calibrated using either a $LaB_6$ or Si NIST standard reference material. A Chebyshev function was used to fit the background for all samples, unless stated otherwise. Both Stephens anisotropic broadening model and pseudo-Voigt functions (TCHZ model) are used to capture peak shapes.[4] Crystallographic information files generated from TOPAS were used in CrystalMaker software for visualization. All XRD plots (and heat maps) in this work use square-root-intensity scaling unless stated otherwise.

All uncertainties are reported using the values generated from TOPAS and are only given for values that were refined independently. Uncertainties from TOPAS reflect the variance and covariance of refined parameters but do not capture sample-to-sample variability. This can be seen in Table S12, where we show the refined orthorhombic distortion parameter, $\delta_O$, for two independent samples (different powder synthesis batches) each of the $Na^+$-vacancy ordered samples ($x$ = 2/3, 1/2, and 1/3 in $Na_xNi_{1/3}Mn_{2/3}O_2$). While the uncertainties from TOPAS are to the fourth or fifth decimal place, the variability between samples is larger. This is especially true for $x$ = 1/3.

We report refinements for the materials in this study using the space groups $P222_1$, *Cmcm*, and *Pnmm*. The choice of each subgroup is described below followed by the transformation matrices to convert each to *Cmcm*. In this work, we refine all XRD data in space group *Cmcm* (in addition to other space groups where appropriate) to be able to compare the magnitude of the orthorhombic distortion across materials.

i. *Refinement for phases with $Na^+$-vacancy ordering*

   a. *$Na_{2/3}Ni_{1/3}Mn_{2/3}O_2$*

Both sXRD and NPD data for $Na_{2/3}Ni_{1/3}Mn_{2/3}O_2$ were co-refined in space group $P222_1$ with atomic coordinates shared between the two datasets. This space group was identified via peak indexing and *ab initio* structure solution in TOPAS, using the charge flipping algorithm.[5,6] The co-refinement allowed lattice parameters and thermal displacements to refine separately between the two data sets to reflect the facts that (1) the data were collected at different (ambient) temperatures and (2) neutron and X-ray scattering cross sections differ (Tables S1-S2, Fig. S1). Partial occupancies on Ni and Mn sites were used to refine the NPD data to reflect the level of stacking faults in the material. However, the total site occupancies were constrained to be one. The global composition was constrained to reflect the ICP-OES result. A model with three Na sites was used with full site occupancy, which does not perfectly capture the $Na^+$-vacancy ordering peaks (see difference line for sXRD data in Fig. S1). This is improved when faults are considered (Fig. 1d of main text and a Fig. S2 for a direct comparison). The co-refined atomic positions were then used to construct a supercell in the $P1$ space group for stacking fault modeling (Table S3).

For constructing the supercell for stacking fault modeling, Ni and Mn dominant sites were set to be fully occupied by Ni and Mn, respectively. Atomic coordinates were fixed while lattice parameters and thermal displacements were allowed to be refined. Three different stacking vectors ($\mathbf{t}_1$, $\mathbf{t}_2$, and $\mathbf{t}_3$) were considered as described in the main text, using notation borrowed from previous literature.[5] The relationship between these vectors and the metal stacking faults are shown in Figure S3. Several different combinations of these fault vectors were tested, as seen in Table S4. Simulated sequences of $N_V$ = 50 stacking vectors were generated 200 times during the refinement to capture the average effects of the three stacking vectors over a whole crystallite. Note that the fault model was only applied to NPD data, since this can distinguish between Ni and Mn sites.

*b. $Na_{1/2}Ni_{1/3}Mn_{2/3}O_2$*

The refinement for ordered $Na_{1/2}Ni_{1/3}Mn_{2/3}O_2$ used a starting model from previous literature on $Na_{1/2}CoO_2$ in *Pnmm*.[6] This model cannot account for Ni/Mn honeycomb ordering but is sufficient for describing the sXRD data presented in this work due to the similar scattering factors of Ni and Mn. Na and O positions along the [100] direction were refined along with the lattice parameters and peak shape functions (Figure S5 and Table S11). Full occupancies of $Na^+$ positions are used to reflect the low amount of faults present in this sample and the accuracy of the simple row ordering model of $Na_f$ and $Na_e$ sites. We note *Pnmm* is not a subgroup of the high-symmetry hexagonal phase of NNM, $P6_3/mmc$, found in the disordered (undistorted) states ($x = 0.59$, $0.42$). As such, a lattice in the *Pnmm* space group cannot undergo a second-order phase transition to one in the $P6_3/mmc$ space group.

*ii. Refinements in the Cmcm space group*

*a. Ambient temperature XRD*

Our refinements of data in main text Figures 2-4 used the P2 structural model in the *Cmcm* space group because it provides a common set of lattice parameters by which the orthorhombic distortion parameter ($\delta_o$) could be calculated. ISODISTORT software[7] was used to generate the orthorhombic *Cmcm* cell from the parent $P6_3/mmc$ cell commonly used in literature, which has hexagonal symmetry.[8] This group/subgroup relationship allows for a second-order phase transition as a group/subgroup relationship is a necessary (but not sufficient) condition for a second-order phase transition. We note, however, that the *Cmcm* space group cannot capture $Na^+$-vacancy ordering, and thus it provides an incomplete description of the ordered states of NNM.

For all ordered samples ($x = 2/3$, $1/2$, and $1/3$), the total sodium stoichiometry was fixed based on the expected amount for each phase. For $x = 0.59$ and $0.42$, which are disordered (undistorted), several different parameters were fixed in the refinement to account for their hexagonal symmetry. Specifically, the *b*-parameter is set equal to $\sqrt{3}a$, leading to $\delta_o = 0$. Also, the displacements of all atoms along the [010] direction were fixed to be zero. Note that fixing these parameters is equivalent to refinement in the $P6_3/mmc$ space group, and reasonable fits were achieved in all cases (see Fig. S4 and Tables S5, S7-S10).

*b. Operando XRD*

Sequential Rietveld refinement on the lab-scale XRD data was used to determine the $\delta_o$ values over a narrow range in sodium stoichiometry ($0.625 < x < 2/3$). The orthorhombic distortion was constrained to be greater than or equal to zero, where zero represents refinement in a hexagonal setting. Due to poor signal-to-noise for $x < 0.625$, it was not possible to refine $\delta_o$ over the whole sodium stoichiometry range. Only pseudo-Voigt functions were used to capture peak shapes. The symmetry is considered orthorhombic if $\delta_o$ is greater than or equal to 0.01%, and hexagonal otherwise.

*c. Variable temperature XRD*

Sequential Rietveld refinement was conducted to refine the lattice parameters and hence $\delta_o$ from the variable temperature sXRD data assuming second-order transitions with continually decreasing $\delta_o$. For the first sample in each dataset, both horizontal and vertical capillary displacement corrections were refined in addition to the zero error. These extrinsic experimental parameters were fixed for refinement of subsequent scans. Furthermore, anomalous scattering from the HTK 1200N heater necessitated a user-defined background fit in TOPAS rather than a Chebyshev function, which was fixed based on the first scan in each dataset (see Figs. S11-S12, Tables S13-S14). The total sodium stoichiometry was set to 2/3 and 1/2 for the $Na_{2/3}Ni_{1/3}Mn_{2/3}O_2$ and $Na_{1/2}Ni_{1/3}Mn_{2/3}O_2$ samples, respectively. Sequential refinement was first performed for all scans using hexagonal symmetry ($\delta_o$= 0 and the displacements of all atoms along the [010] direction fixed at zero). Next, the refinement was repeated while allowing these parameters to freely vary. The transition temperature is reported as the temperature at which $\delta_o$ first goes to zero, or less than 0.01%, on heating. For $Na_{1/2}Ni_{1/3}Mn_{2/3}O_2$, reaction with the PVDF binder and carbon black is shown starting around 200 ºC by a decreasing integrated area of the (002) peak (Fig. S16). The fit quality ($R_{wp}$) as a function of temperature can be seen for both $Na_{2/3}Ni_{1/3}Mn_{2/3}O_2$ and $Na_{1/2}Ni_{1/3}Mn_{2/3}O_2$ in Figure S10.

*iii. Transformation matrices for lattice vectors between different space groups*

*a. Cmcm to Pnmm*

$$\begin{pmatrix} a_{Pnmm} \\ b_{Pnmm} \\ c_{Pnmm} \end{pmatrix} = \begin{pmatrix} 0 & 1 & 0 \\ 2 & 0 & 0 \\ 0 & 0 & 1 \end{pmatrix} \begin{pmatrix} a_{Cmcm} \\ b_{Cmcm} \\ c_{Cmcm} \end{pmatrix}$$

*b. Cmcm to P222₁*

$$\begin{pmatrix} a_{P2221} \\ b_{P2221} \\ c_{P2221} \end{pmatrix} = \begin{pmatrix} 3 & 0 & 0 \\ 0 & 1 & 0 \\ 0 & 0 & 1 \end{pmatrix} \begin{pmatrix} a_{Cmcm} \\ b_{Cmcm} \\ c_{Cmcm} \end{pmatrix}$$

*c. $P6_3/mmc$ to Cmcm*

$$\begin{pmatrix} a_{Cmcm} \\ b_{Cmcm} \\ c_{Cmcm} \end{pmatrix} = \begin{pmatrix} 1 & 0 & 0 \\ 1 & 2 & 0 \\ 0 & 0 & 1 \end{pmatrix} \begin{pmatrix} a_{P6_3/mmc} \\ b_{P6_3/mmc} \\ c_{P6_3/mmc} \end{pmatrix}$$

**Supplementary Discussions**

*Supplementary Discussion 1: Metal Stacking Faults in* $Na_{1/2}Ni_{1/3}Mn_{2/3}O_2$

In the main text, we state that the $x = 1/2$ phase of NNM likely has fewer transition-metal stacking faults compared to the $x = 2/3$ phase. Although transition-metal stacking faults were identified directly by NPD only for the $x = 2/3$ phase, sXRD can also provide indirect signatures of such disorder. In layered oxides, metal stacking faults perturb the local arrangement of the transition metal slabs and the coupled interlayer $Na^+$-vacancy ordering, leading to anisotropic peak broadening that is observable in sXRD patterns (Figure S6).

To assess whether signatures of metal stacking faults vary among compositions using sXRD data, we compare Rietveld refinements in space group *Cmcm* with and without Stephens anisotropic broadening parameters. Stephens parameters can significantly improve Rietveld refinements when anisotropic strain broadening or other microstructural defects (such as stacking faults) are not explicitly included in the structural model.[4] We refine three of the *ex situ* samples, $x = 2/3$, 0.59, and 1/2 using just pseudo-Voigt peak shapes (no Stephens parameters). Since such refinements have fewer free parameters, we expect the $R_{wp}$ values to increase when the Stephens model is not included. Indeed, the $R_{wp}$ values without (with) the Stephens terms are 9.71% (4.46%), 5.65% (3.45%), and 6.27% (4.94%) for $x = 2/3$, 0.59, and 1/2, respectively. The increase in $R_{wp}$ upon removing the Stephens model is largest for $x = 2/3$. In Figure S6, we show the refinements without the Stephens model. The difference curve at $x = 2/3$ exhibits pronounced tails under reflections with layering character (highlighted peaks in Fig. S6) that are not captured by pseudo-Voigt peak shapes. Similar tails have previously been associated with stacking disorder in layered oxides.[9,10] Together with the NPD results demonstrating metal stacking faults at $x = 2/3$, these observations indicate that the signatures of stacking disorder are substantially weaker at $x = 0.59$ and $x = 1/2$.

The increase in electrostatic repulsion between transition metal layers in NNM upon desodiation (manifesting in a larger *c* lattice parameter) can in principle be mitigated by reducing stacking arrangements that have vertically stacked $Mn^{4+}$-$Mn^{4+}$ pairs. By inspection of the stacking vectors included in our model for the NPD data analysis (Fig. S3), we see that the incidence of these $Mn^{4+}$-$Mn^{4+}$ pairs across multiple unit cells is largest for the **t2** vector and smallest for the parent **t1** vector. Thus, when considering these three vectors, the **t1** vector is likely to have the lowest energy, which may explain the relatively fewer stacking faults for $x$ = 1/2 compared to $x$ = 2/3. We also note that our model does not consider all possible stacking vectors due to computational modeling cost (see Refinement Details above). It is plausible that other stacking sequences will contribute to a yet more complex energy landscape. Nevertheless, our model adequately describes the peak broadening observed in our NPD data at $x$ = 2/3 and, thus, the total fraction of stacking faults in NNM.

*Supplementary Discussion 2: Negligible Orthorhombic Distortion in* $Na_{1/3}Ni_{1/3}Mn_{2/3}O_2$

As mentioned in the main text, severe anisotropic peak broadening and overlap of the D2 peaks with the (008) peak added difficulty to the refinement of the $x$ = 1/3 phase of NNM. While Table S12 shows that both repeats at $x$ = 1/3 have a positive orthorhombic distortion, the values are small (<0.06%), and the sample-to-sample variation is relatively large (~0.05%). Thus, the refined distortion for $x$ = 1/3 may be attributable to overfitting when using orthorhombic symmetry, which has more free parameters than the hexagonal case. This is reflected in Figure S7, where we show the differences in fit quality ($R_{wp}$) when refining using either orthorhombic or hexagonal symmetry, here using the *Cmcm* or $P6_3/mmc$ space group, respectively. With hexagonal symmetry, $\delta_o$ is fixed at zero, and all atoms sit at high symmetry sites (no displacement). The fits for $x$ = 2/3 and 1/2 experience significant improvement when using an orthorhombic unit cell instead of a hexagonal one. For $x$ = 0.59, 0.42, and 1/3, the improvement is marginal. Because of the small distortion value, large sample-to-sample variation, and marginal improvement in fit quality when using orthorhombic versus hexagonal unit cell symmetry in the $x$ = 1/3 sample, we conclude that it has a negligible orthorhombic distortion.

## Supplementary Figures

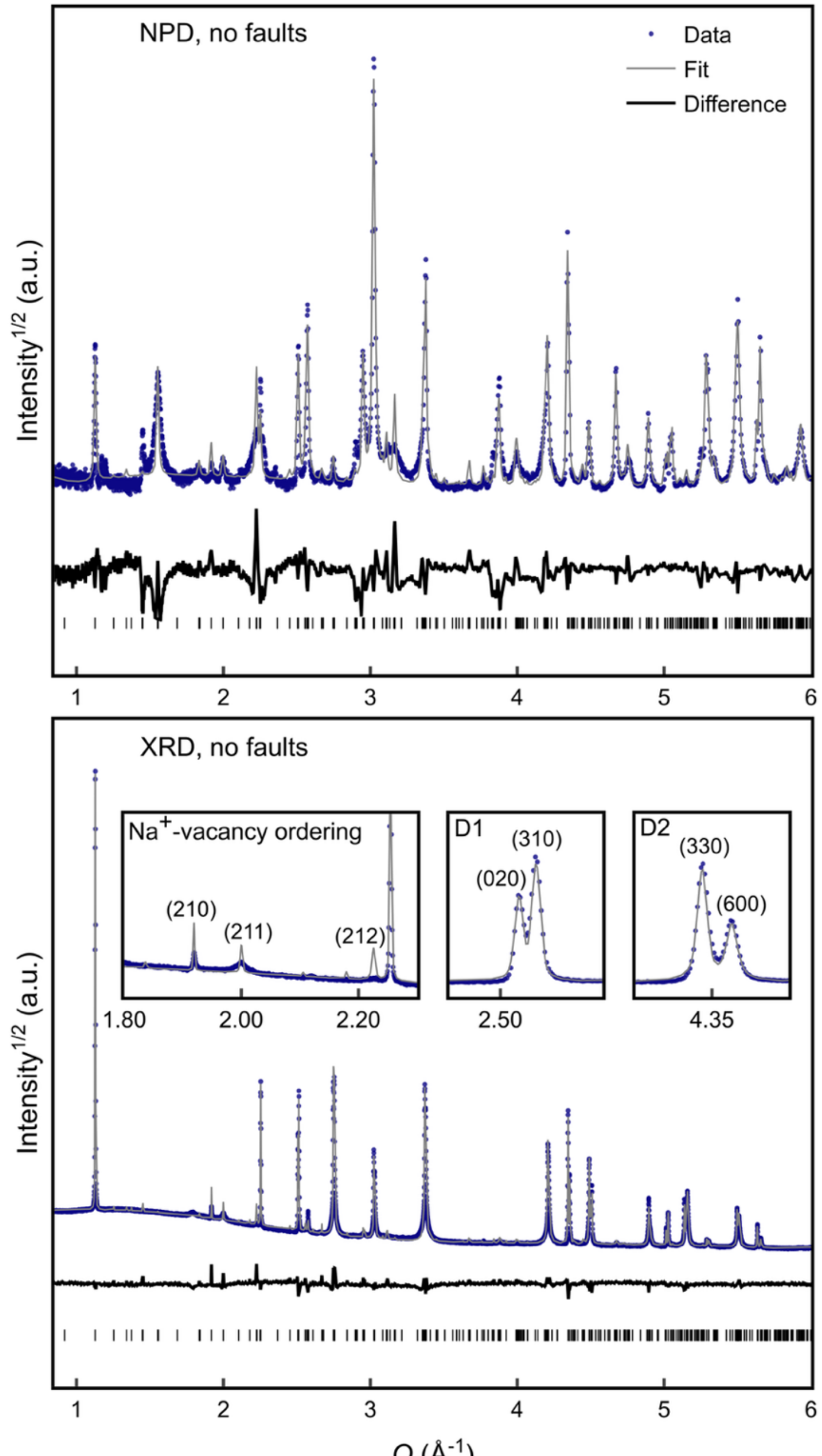

**Figure S1.** Co-refinement of NPD data (top) and sXRD data (bottom) for $Na_{2/3}Ni_{1/3}Mn_{2/3}O_2$ without stacking faults in space group $P222_1$. The fit for the NPD is poor ($R_{wp}$ = 11.76%) while that for the sXRD is acceptable ($R_{wp}$ = 4.65%), indicating significant transition-metal stacking faults are present. Insets show $Na^+$-vacancy ordering peaks and D1 and D2 sets of peaks discussed in the main text. Tick marks denote calculated reflections. See Tables S1 and S2 for full refinement details. $Q$ is the magnitude of the scattering wavevector. The neutron wavelength was 1.5 Å and the X-ray wavelength was 0.7296 Å.

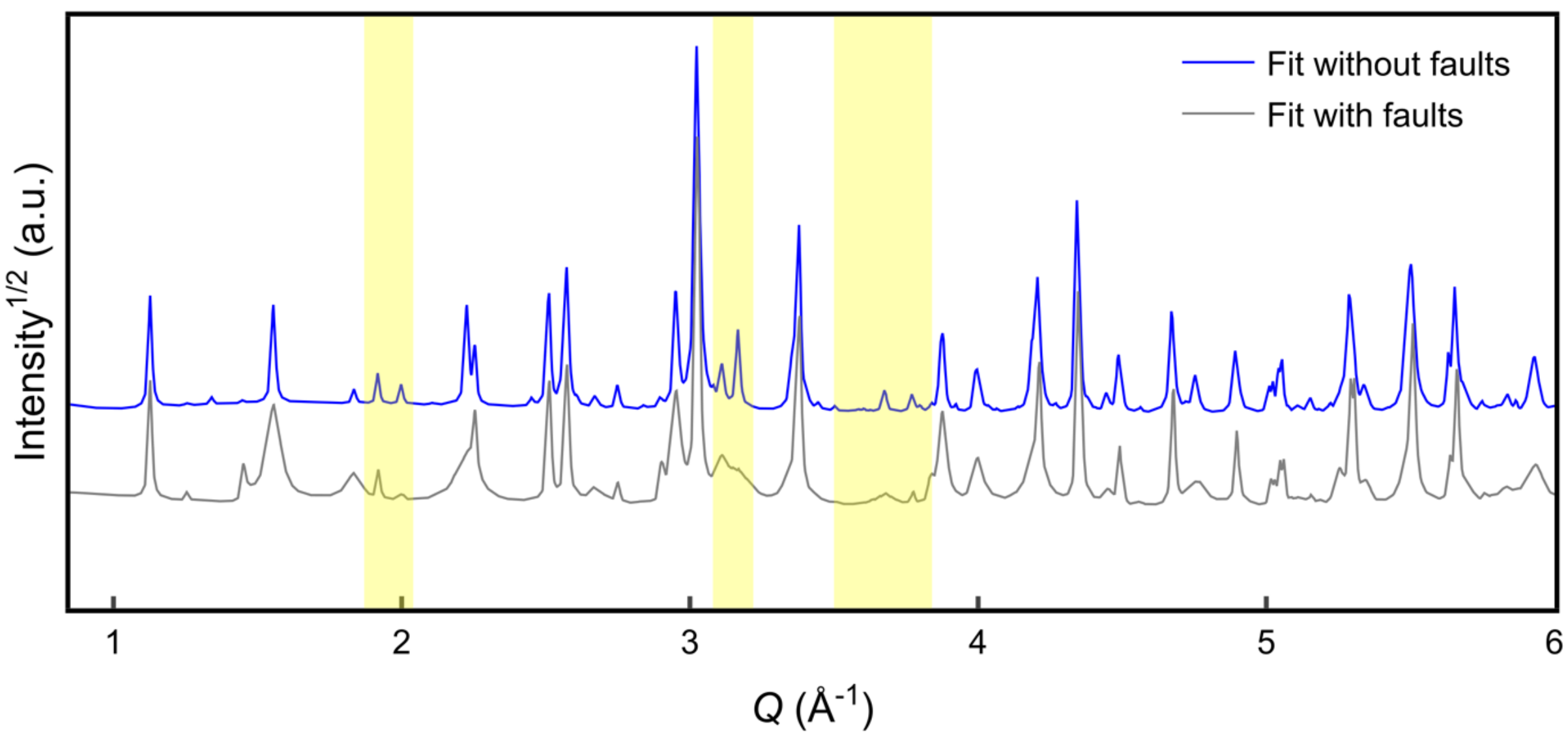

**Figure S2.** Direct comparison of the influence of our NPD stacking fault model (Table S3) on $Na^+$-vacancy superlattice reflections (highlighted regions) compared to the fault-free case (Table S2).

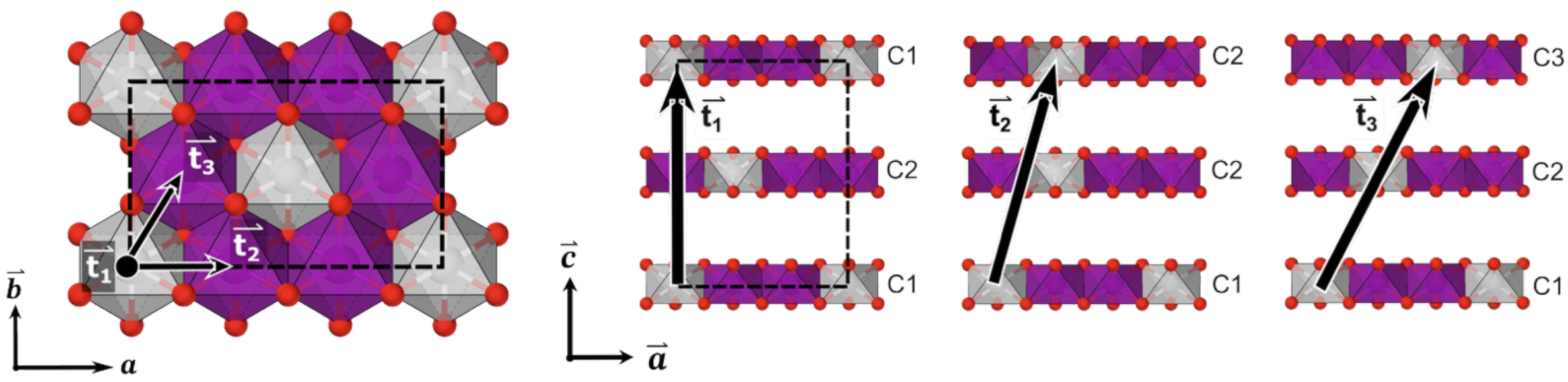

**Figure S3.** Relationship between stacking vectors $\mathbf{t_1}$, $\mathbf{t_2}$, and $\mathbf{t_3}$ in the $P222_1$ unit cell (black dashed boxes) and the types of transition-metal stacking faults (no faults, $C_2C_2$, and $C_1C_2C_3$). Mn and Ni octahedra are shown in purple and gray, respectively, with O atoms in red. Na atoms are omitted for clarity.

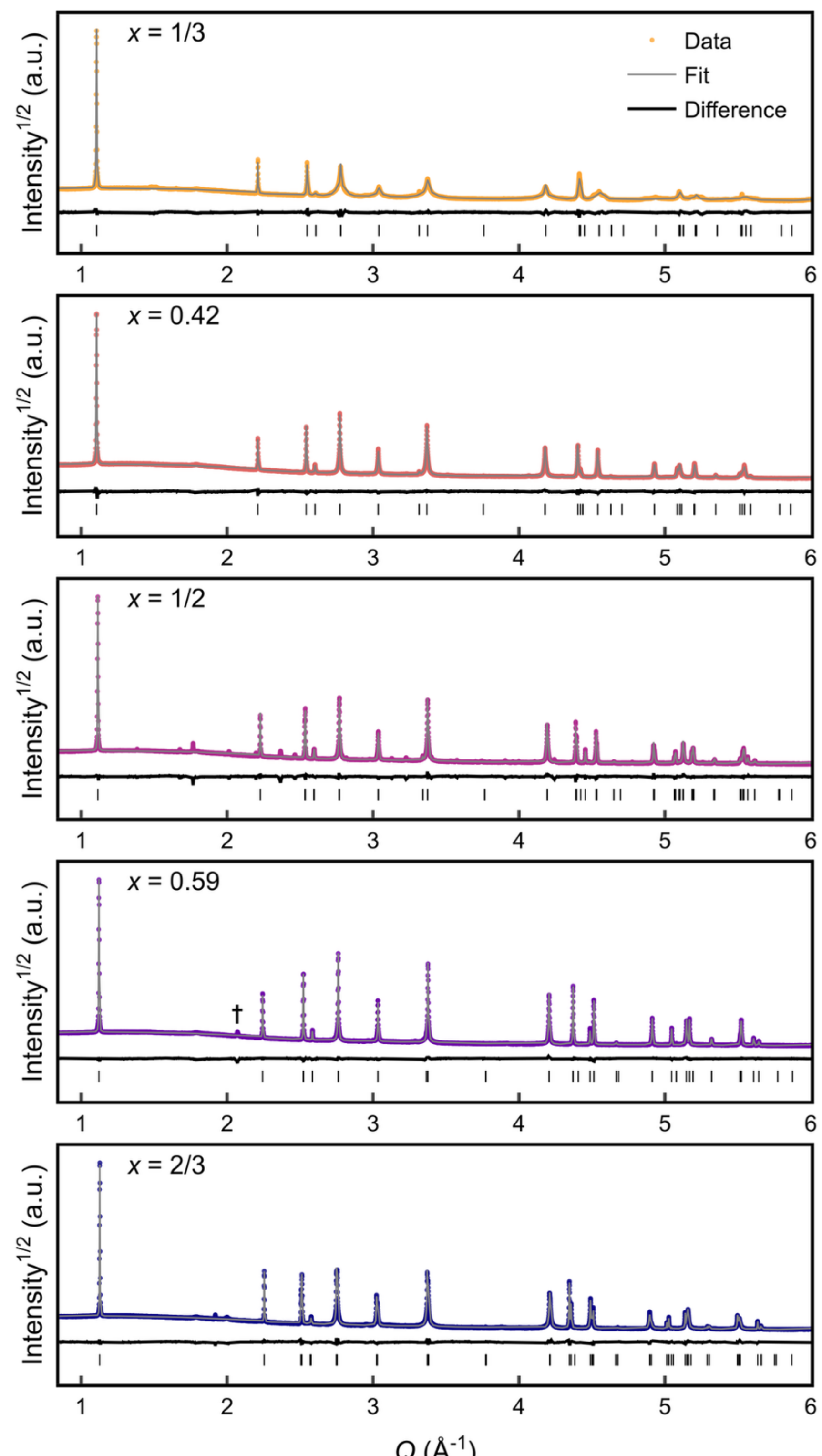

**Figure S4.** Rietveld refinements in the *Cmcm* space group for $x$ = 2/3, 0.59, 1/2, 0.42, and 1/3 in $Na_xNi_{1/3}Mn_{2/3}O_2$. Dagger in the pattern for $x$ = 0.59 indicates impurity from cotton wool, which was used to help pack powder into the capillary. Tick marks denote calculated reflections. See Tables S5, S7-S10 for Rietveld refinement parameters. The X-ray wavelength was 0.7296 Å.

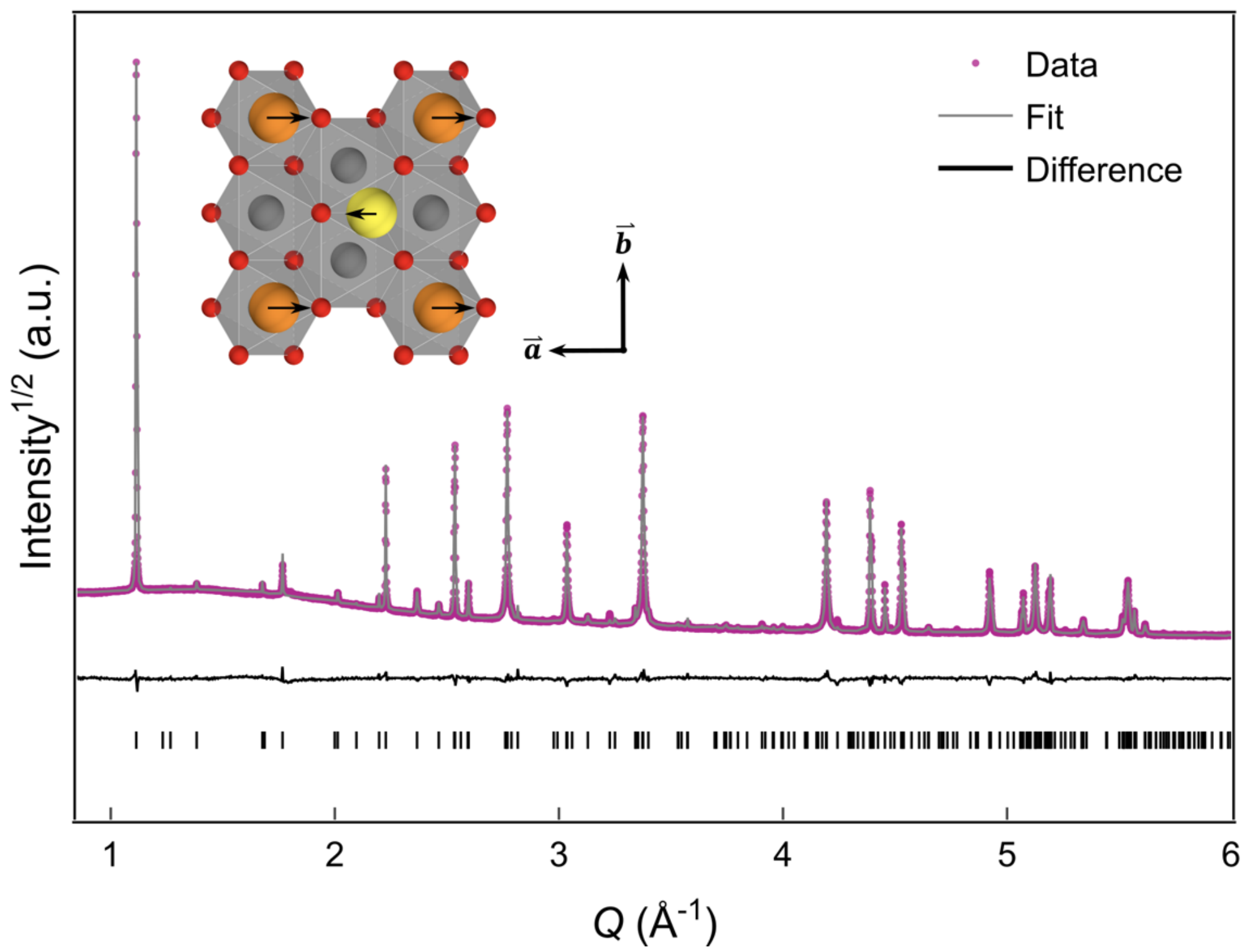

**Figure S5.** Rietveld refinement for the sXRD pattern of $Na_{1/2}Ni_{1/3}Mn_{2/3}O_2$ in space group *Pnmm*. Transition metal octahedra are shown in gray (since this model does not distinguish between Ni and Mn sites), while $Na_f$ and $Na_e$ atoms are in orange and yellow, respectively. O atoms are in red. Arrows indicate displacement directions of $Na_f$ and $Na_e$ sites away from high symmetry positions and are not to scale. Note the *a* and *b* directions in the *Pnmm* model are rotated 90° with respect to their *Cmcm* counterparts (see transformation matrices given in Refinement Details Section *i*). Tick marks denote calculated reflections. See Table S11 for full refinement details. The wavelength was 0.7296 Å.

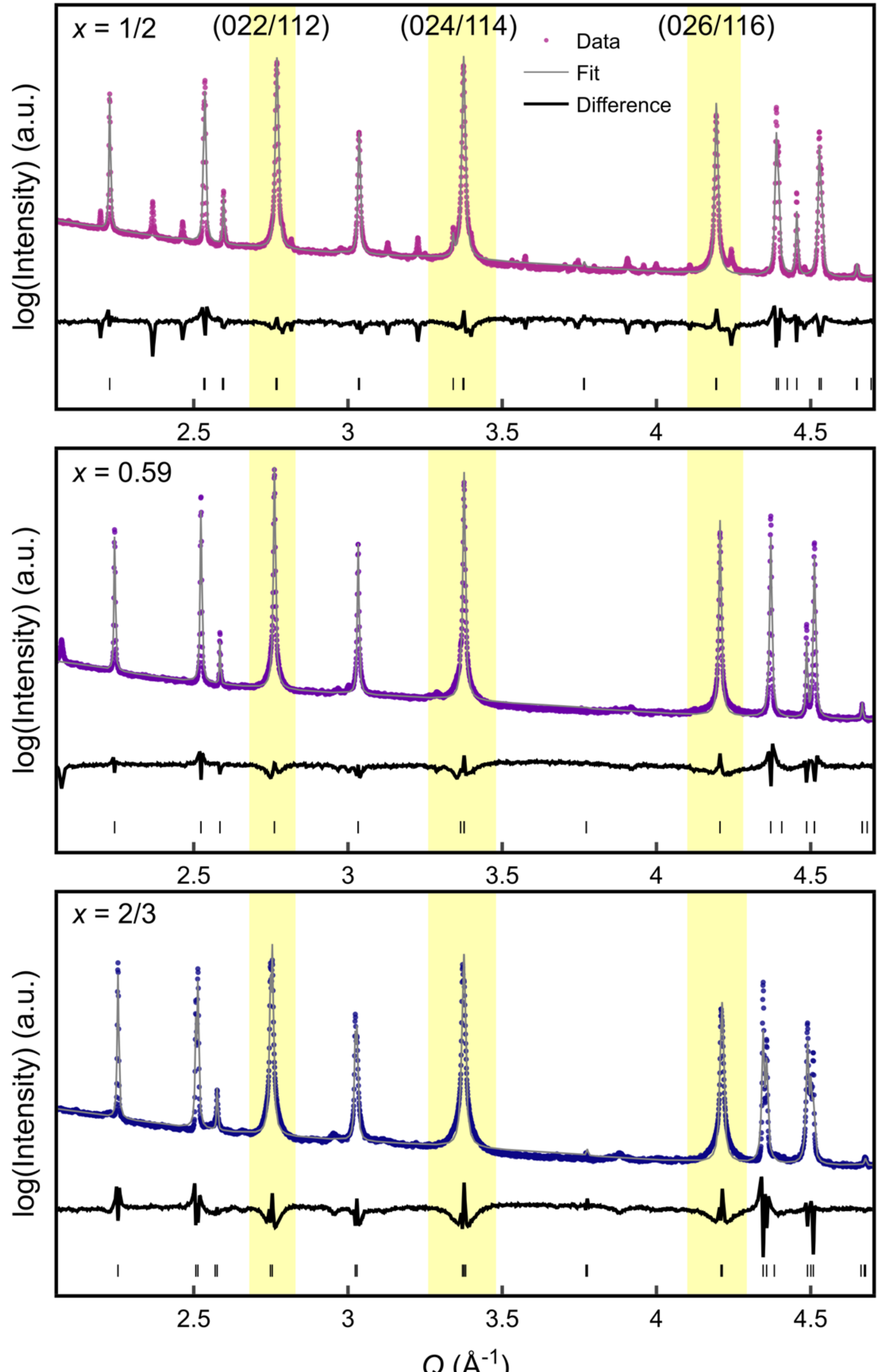

**Figure S6.** Rietveld refinements in the *Cmcm* space group for $x = 2/3$, 0.59, and 1/2 using only pseudo-Voigt peak shape parameters. Highlighted regions show Bragg peaks with layering character. The difference curve at $x = 2/3$ has the largest residuals (likely due to stacking faults) while the differences curves for $x = 0.59$ and 1/2 have similar residuals in these regions. $R_{wp}(x = 2/3) = 9.71\%$; $R_{wp}(x = 0.59) = 5.65\%$; $R_{wp}(x = 1/2) = 6.27\%$.

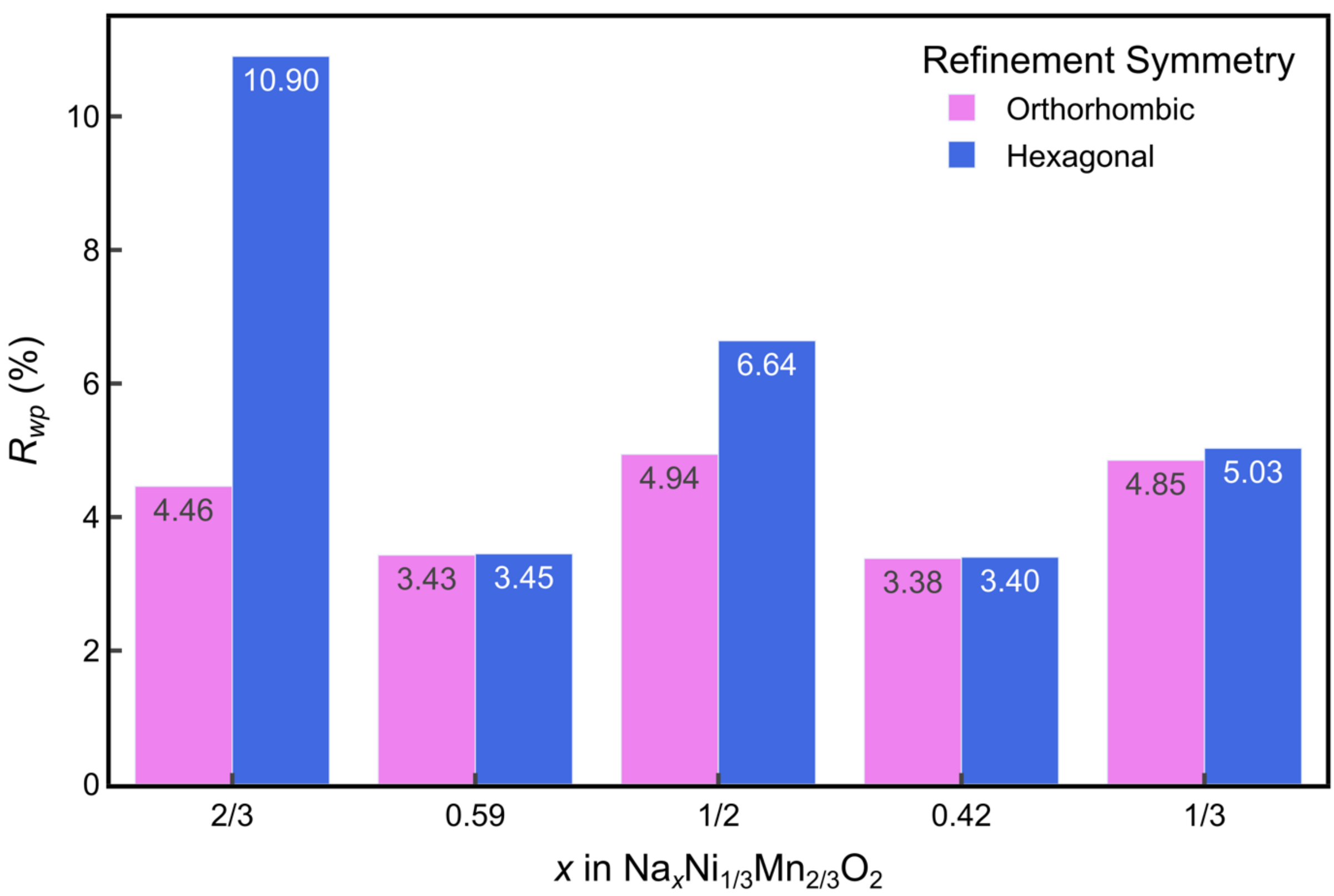

**Figure S7.** Improvement in fit quality by lowering unit cell symmetry from hexagonal ($P6_3/mmc$) to orthorhombic ($Cmcm$) for harvested samples from Figure 2 in the main text. The fits for each sample are better with orthorhombic symmetry, as expected. However, for $x$ = 0.59, 0.42, and 1/3, the improvement in fit quality is marginal when lowering the symmetry from hexagonal to orthorhombic.

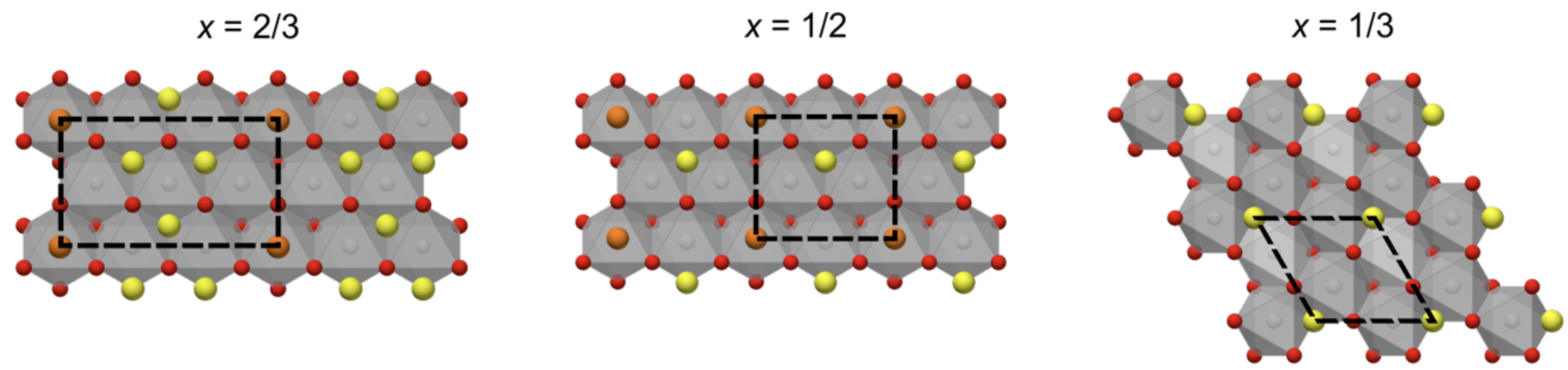

**Figure S8.** $Na^+$-vacancy ordering models for different ordered phases in $Na_xNi_{1/3}Mn_{2/3}O_2$ (with $x$ = 1/2 and 1/3 models reported in previous literature).[6,11] Black dashed boxes show the in-plane repeat units for the $Na^+$-vacancy ordering, which is orthorhombic for $x$ = 2/3 and 1/2. Transition metal octahedra are shown in gray (not distinguishing between Ni and Mn sites here, for simplicity), while $Na_f$ and $Na_e$ atoms are in orange and yellow, respectively. O atoms are in red. $Na^+$ displacements have been set to zero for simplicity.

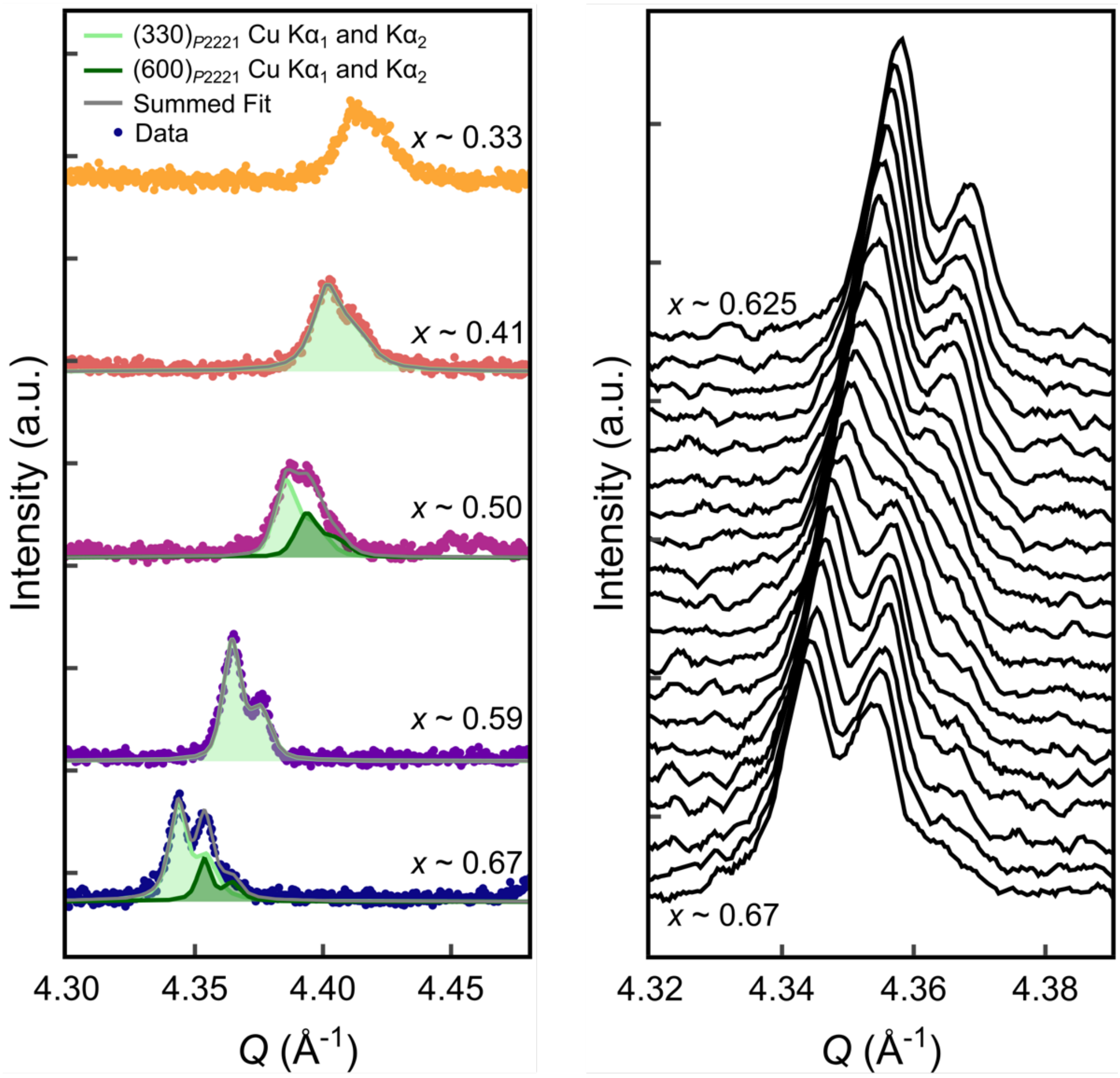

**Figure S9**. Analysis of *operando* XRD data. Left plot shows XRD patterns from the waterfall plot in main text Fig. 3a with individual peak fits shown for visualization purposes. When the ratio between the two peaks is not 2:1 (the intensity ratio for Cu $K\alpha_1$: $K\alpha_2$), an orthorhombic distortion is present, as can be seen with our individual peak fits (green). Note that due to poor signal-to-noise and the overlap of the (008) peak with the D2 peaks, we did not attempt individual peak fitting for $x = 1/3$ here. Right plot shows evolution of operando XRD patterns over $0.625 < x < 2/3$ in $Na_xNi_{1/3}Mn_{2/3}O_2$, showing a continuous change in symmetry consistent with second-order behavior. The data in the right plot has been re-binned to a step size of 0.02º for clarity.

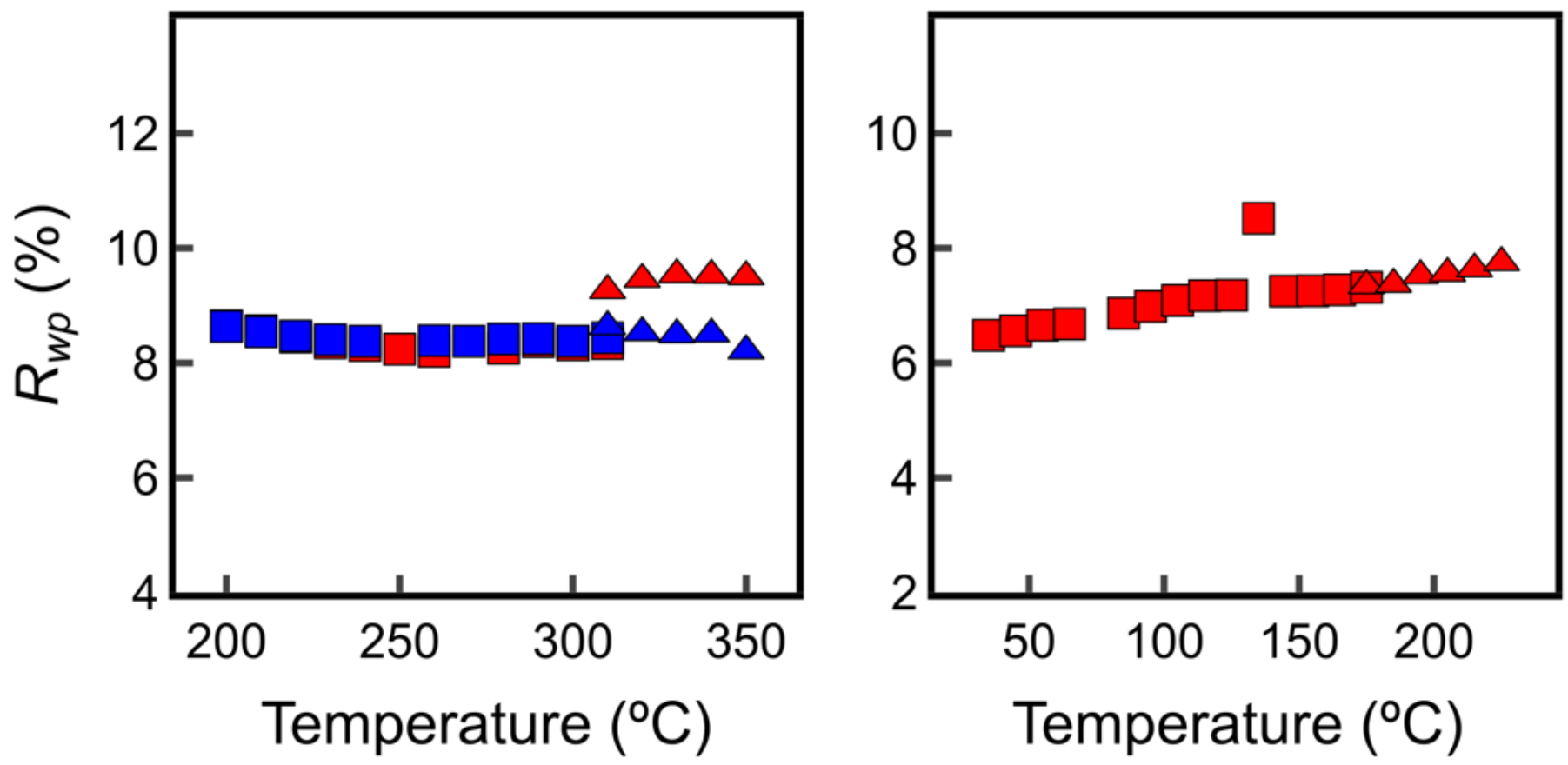

**Figure S10**. Fit quality as a function of temperature for $Na_{2/3}Ni_{1/3}Mn_{2/3}O_2$ (left) and $Na_{1/2}Ni_{1/3}Mn_{2/3}O_2$ (right). Red and blue colors represent heating and cooling traces, respectively. Squares and triangles represent orthorhombic and hexagonal symmetry being used for the refinement. Two points have been removed from the left plot for clarity due to a detector issue that did not affect accurate refinement of lattice parameters.

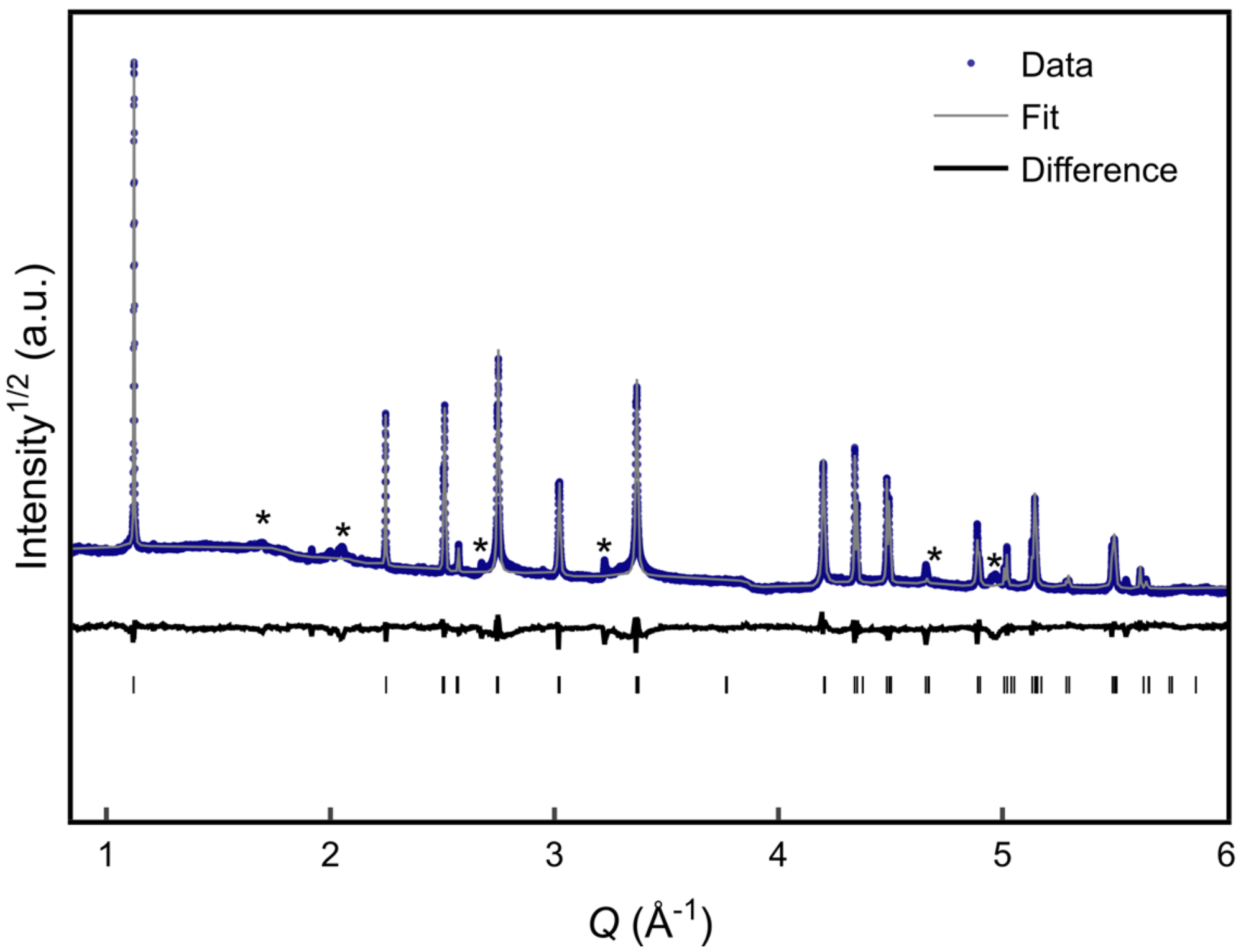

**Figure S11.** sXRD pattern for $Na_{2/3}Ni_{1/3}Mn_{2/3}O_2$ at 200 ºC, refined in space group *Cmcm*. Asterisks denote scattering from the Anton Paar HTK 1200N heating stage. Tick marks denote calculated reflections. See Table S13 for full refinement details.

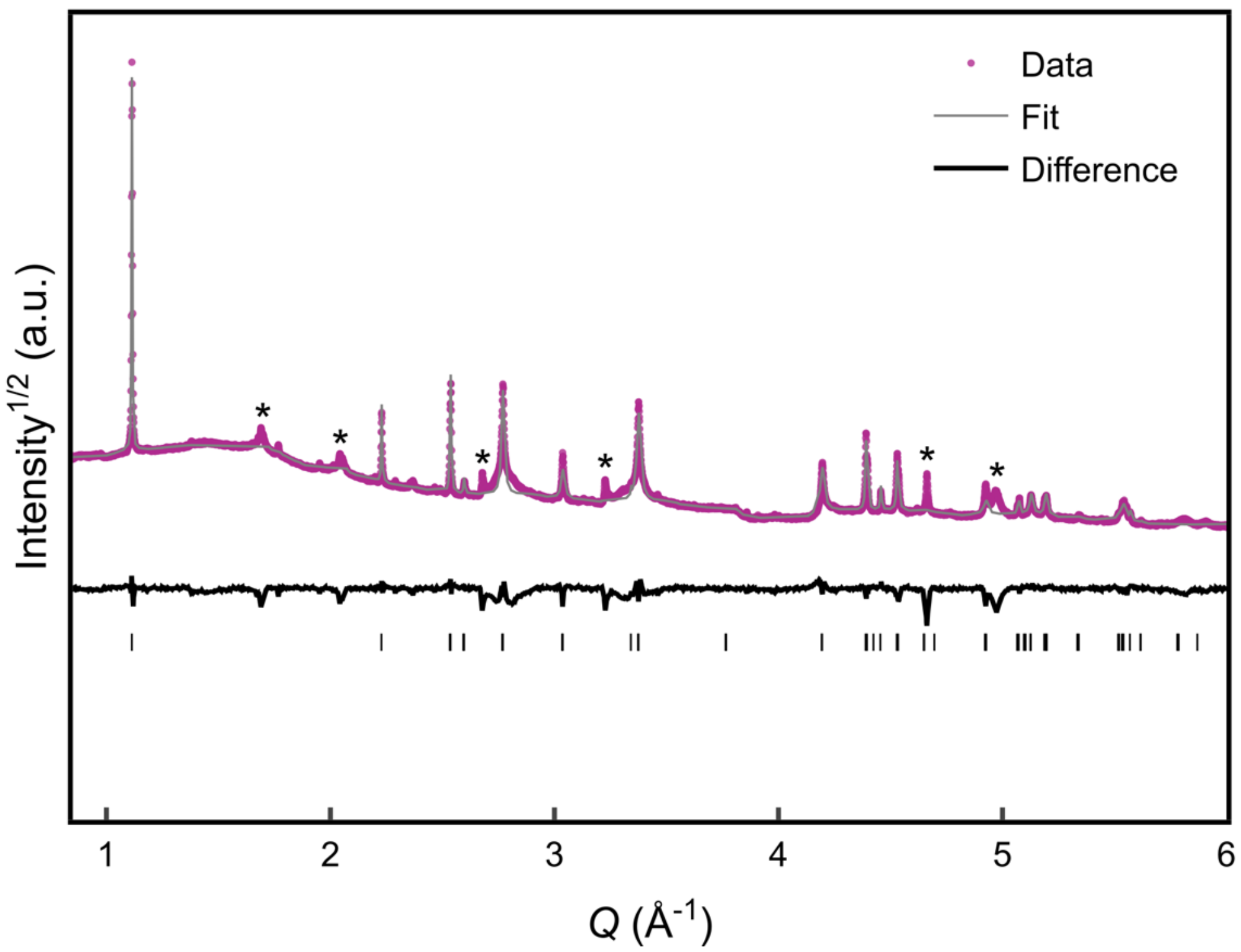

**Figure S12.** sXRD pattern for $Na_{1/2}Ni_{1/3}Mn_{2/3}O_2$ at 35 ºC, refined in space group *Cmcm*. Asterisks denote same peaks from the Anton Paar heater as in Figure S11. Tick marks denote calculated reflections. See Table S14 for full refinement details.

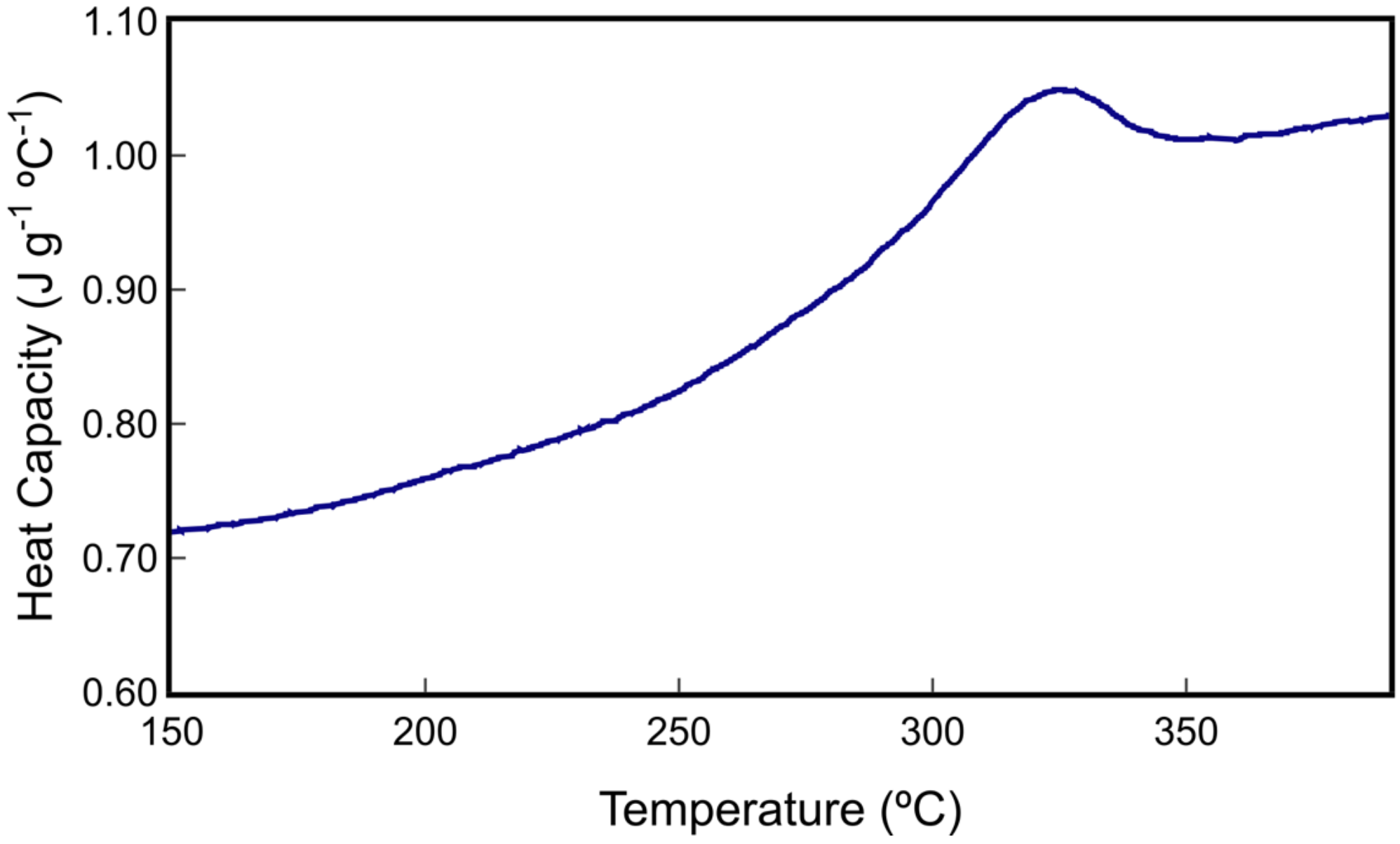

**Figure S13.** DSC scan for $Na_{2/3}Ni_{1/3}Mn_{2/3}O_2$. The broad peak in heat capacity between ~250 ºC and 340 ºC indicates a second-order phase transition is likely occurring in this region.

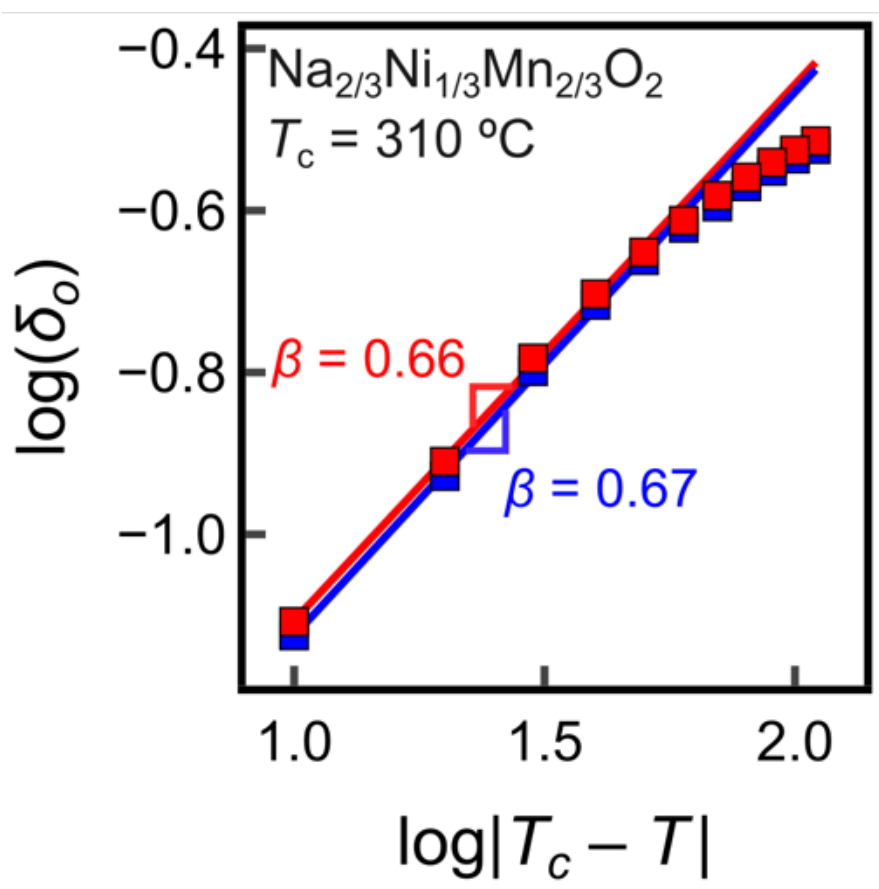

**Figure S14**. Critical exponent ($\beta$) for the temperature-induced second-order phase transition at $Na_{2/3}Ni_{1/3}Mn_{2/3}O_2$. The critical exponent was fit based on a simple power law equation: $\delta_o = A|T_c - T|^\beta$ where $\delta_o$ is the orthorhombic distortion parameter defined in the main text, $A$ is a fitting constant, $T_c$ is the critical temperature, and $T$ is the measured temperature. The critical temperature was the first point at which $\delta_o$ went to zero on heating as described in the "Refinement Details" section. Both the heating (red) and cooling (blue) traces are shown. The critical exponent was only fit to the linear portion of this curve, as the linear trend breaks far from the critical point.

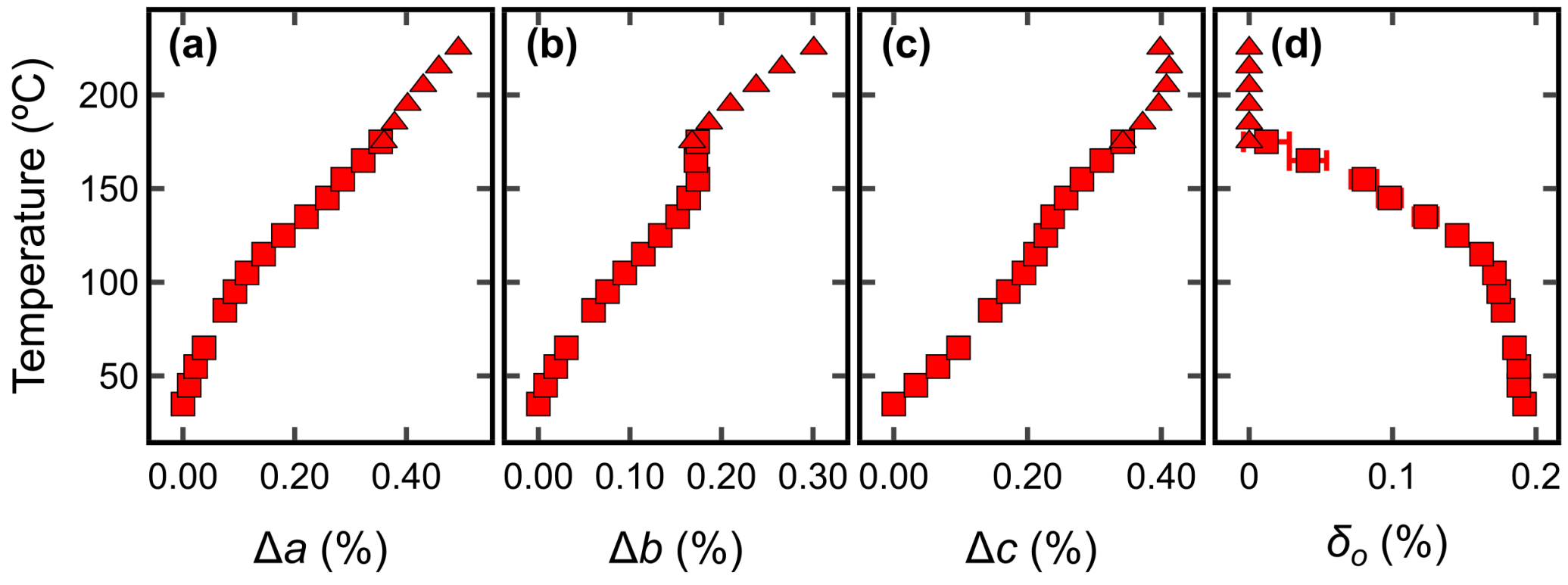

**Figure S15.** Refinement of *in situ* variable temperature sXRD at $x = 1/2$ (main text Figure 5) assuming a continuous phase transition. (a)–(c) Temperature-dependent percentage change (Δ) of *a*-, *b*-, and *c*-parameters of the *Cmcm* unit cell with respect to the first measured values, respectively. (d) Evolution of the orthorhombic distortion parameter ($\delta_o$) with temperature. Error bars are smaller than the size of the markers, except in (d). Squares indicate orthorhombic symmetry was used in the refinement, while triangles indicate hexagonal symmetry was used.

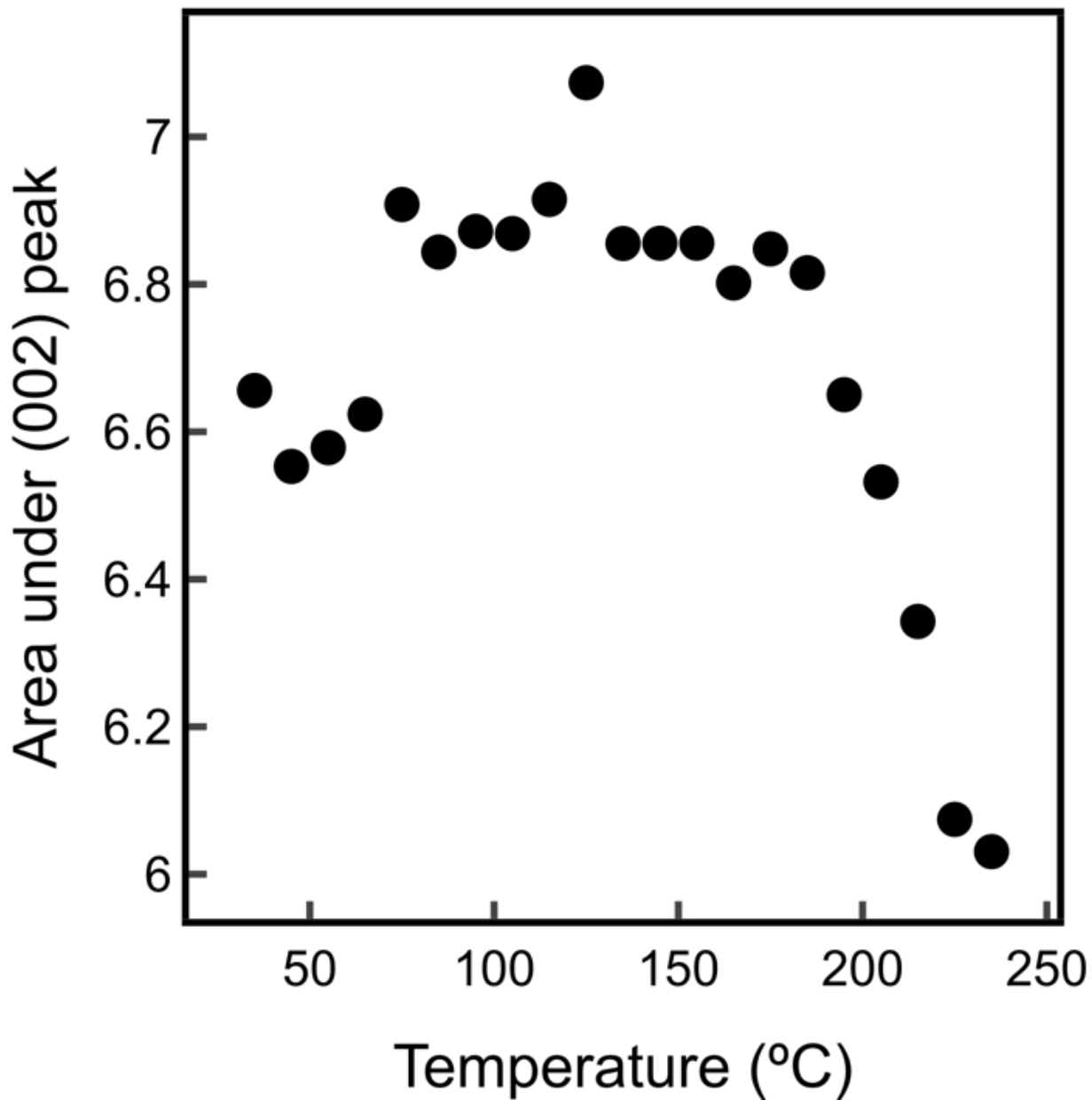

**Figure S16.** Integrated area under the (002) peak for the *in situ* heating sXRD data of $Na_{1/2}Ni_{1/3}Mn_{2/3}O_2$. The drop in integrated area starting near 200 °C likely indicates reaction with PVDF binder (melting point ~177 °C) or carbon black present in the capillary. Carbothermal reduction is not expected below 500 °C.[12]

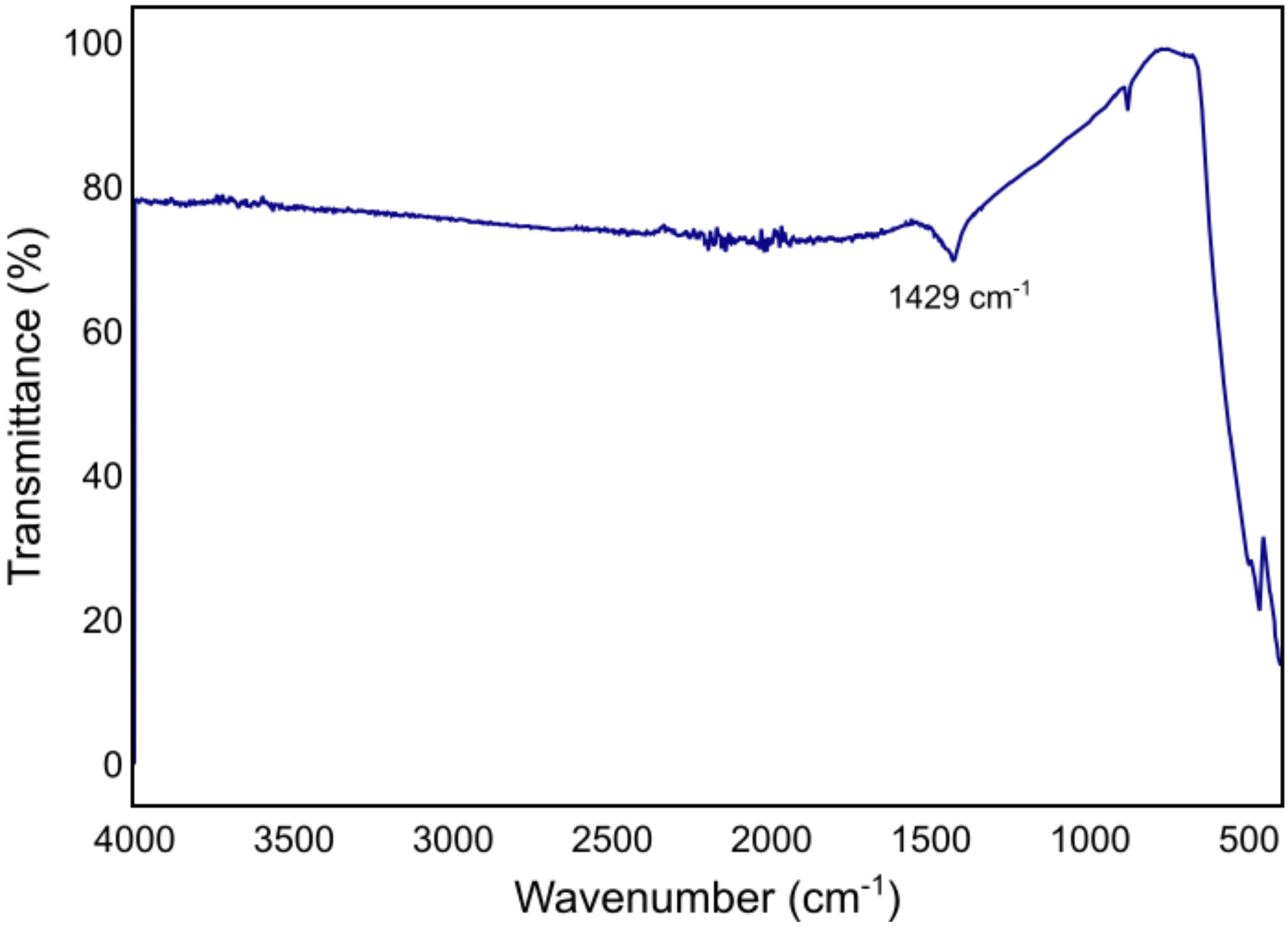

**Figure S17.** FTIR spectrum for $Na_{2/3}Ni_{1/3}Mn_{2/3}O_2$ showing an absorbance peak at 1429 $cm^{-1}$, which indicates residual carbonate species (such as $Na_2CO_3$ and $NaHCO_3$) are present in the sample.

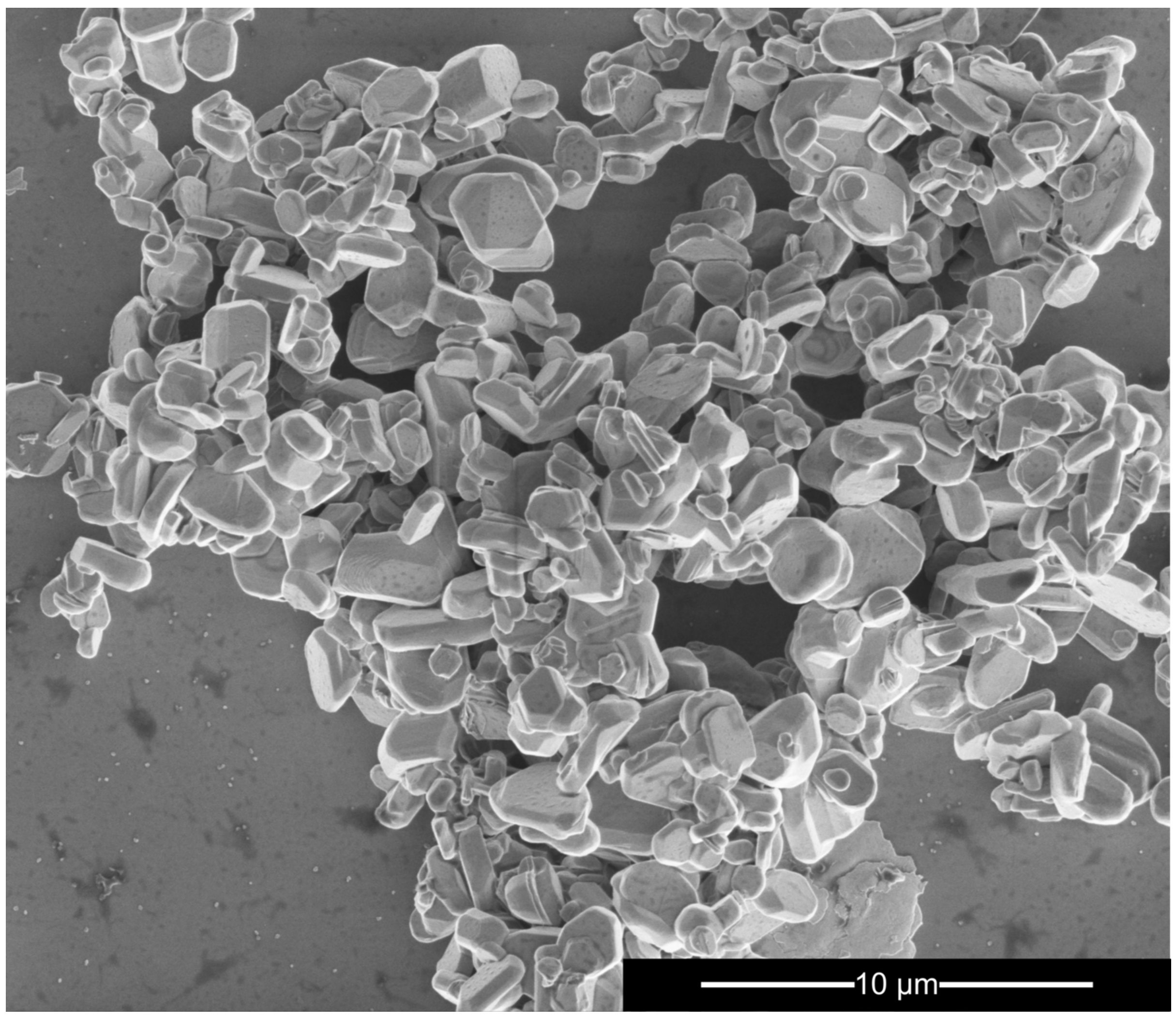

**Figure S18.** Scanning electron micrograph of $Na_{2/3}Ni_{1/3}Mn_{2/3}O_2$ powder showing its platelet-like morphology and size (~1-5 µm primary particles).

## Supplementary Tables

**Table S1.** sXRD refinement results for $Na_{2/3}Ni_{1/3}Mn_{2/3}O_2$ in space group $P222_1$. Ni and Mn occupancies were fixed to those from the NPD refinement in Table S2.

| $a$ (Å) | 8.65208(19) | | | | |
|---|---|---|---|---|---|
| $b$ (Å) | 5.01303(10) | | | | |
| $c$ (Å) | 11.1494(3) | | | | |
| $V$ (Å$^3$) | 483.583(18) | | | | |
| $Z$ | 12 | | | | |
| $R_{wp}$ | 4.65% | | | | |
| Atom | $x$ | $y$ | $z$ | occupancy | $B_{eq}$ (Å$^2$) |
| Na1 | 0 | 0 | 0.25 | 1 | 3.6(3) |
| Na2 | 0.6993(10) | 0.3557(19) | 0.2455(13) | 1 | $= B_{Na1}$ |
| Na3 | 0.5 | 0.870(3) | 0.25 | 1 | $= B_{Na1}$ |
| Ni1 | 0.3396(10) | 0 | 0 | 0.697 | 1.0(4) |
| Ni2 | 0.8402(11) | 0.5 | 0 | 0.679 | $= B_{Ni1}$ |
| Ni3 | 0.1650(19) | 0.5 | 0 | 0.109 | 0.7(2) |
| Ni4 | 0.6667(15) | 0 | 0 | 0.025 | $= B_{Ni3}$ |
| Ni5 | 0 | 0 | 0 | 0.274 | $= B_{Ni3}$ |
| Ni6 | 0.5 | 0.5 | 0 | 0.215 | $= B_{Ni3}$ |
| Mn1 | $= x_{Ni1}$ | 0 | 0 | 0.303 | $= B_{Ni1}$ |
| Mn2 | $= x_{Ni2}$ | 0.5 | 0 | 0.321 | $= B_{Ni1}$ |
| Mn3 | $= x_{Ni3}$ | 0.5 | 0 | 0.891 | $= B_{Ni3}$ |
| Mn4 | $= x_{Ni4}$ | 0 | 0 | 0.975 | $= B_{Ni3}$ |
| Mn5 | $= x_{Ni5}$ | 0 | 0 | 0.726 | $= B_{Ni3}$ |
| Mn6 | $= x_{Ni6}$ | 0.5 | 0 | 0.785 | $= B_{Ni3}$ |
| O1 | 0.0110(7) | 0.6838(10) | 0.0950(7) | 1 | 0.30(10) |
| O2 | 0.3375(6) | 0.6536(10) | 0.1009(5) | 1 | $= B_{O1}$ |
| O3 | 0.6541(8) | 0.6841(10) | 0.0904(6) | 1 | $= B_{O1}$ |
| O4 | 0.1521(8) | 0.1663(11) | 0.0919(5) | 1 | $= B_{O1}$ |
| O5 | 0.5227(7) | 0.1726(10) | 0.0906(6) | 1 | $= B_{O1}$ |
| O6 | 0.8450(7) | 0.1406(10) | 0.0813(5) | 1 | $= B_{O1}$ |

**Table S2.** NPD refinement results for $Na_{2/3}Ni_{1/3}Mn_{2/3}O_2$ in space group $P222_1$ (no stacking faults). Atomic coordinates were co-refined with the sXRD data in Table S1.

| | | | | | |
|---|---|---|---|---|---|
| *a* (Å) | 8.66023(13) | | | | |
| *b* (Å) | 5.01686(6) | | | | |
| *c* (Å) | 11.15582(15) | | | | |
| *V* (Å$^3$) | 484.689(11) | | | | |
| *Z* | 12 | | | | |
| $R_{wp}$ | 11.76% | | | | |
| Atom | *x* | *y* | *z* | occupancy | $B_{eq}$ (Å$^2$) |
| Na1 | 0 | 0 | 0.25 | 1 | 2.79(10) |
| Na2 | 0.6993(10) | 0.3557(19) | 0.2455(13) | 1 | = $B_{Na1}$ |
| Na3 | 0.5 | 0.870(3) | 0.25 | 1 | = $B_{Na1}$ |
| Ni1 | 0.3396(10) | 0 | 0 | 0.697(9) | 0.001(37) |
| Ni2 | 0.8402(11) | 0.5 | 0 | 0.679(9) | = $B_{Ni1}$ |
| Ni3 | 0.1650(19) | 0.5 | 0 | 0.109(9) | 0.001(56) |
| Ni4 | 0.6667(15) | 0 | 0 | 0.025* | = $B_{Ni3}$ |
| Ni5 | 0 | 0 | 0 | 0.274(7) | = $B_{Ni3}$ |
| Ni6 | 0.5 | 0.5 | 0 | 0.215(8) | = $B_{Ni3}$ |
| Mn1 | = $x_{Ni1}$ | 0 | 0 | = 1 – $occ_{Ni1}$ | = $B_{Ni1}$ |
| Mn2 | = $x_{Ni2}$ | 0.5 | 0 | = 1 – $occ_{Ni2}$ | = $B_{Ni1}$ |
| Mn3 | = $x_{Ni3}$ | 0.5 | 0 | = 1 – $occ_{Ni3}$ | = $B_{Ni3}$ |
| Mn4 | = $x_{Ni4}$ | 0 | 0 | = 1 – $occ_{Ni4}$ | = $B_{Ni3}$ |
| Mn5 | = $x_{Ni5}$ | 0 | 0 | = 1 – $occ_{Ni5}$ | = $B_{Ni3}$ |
| Mn6 | = $x_{Ni6}$ | 0.5 | 0 | = 1 – $occ_{Ni6}$ | = $B_{Ni3}$ |
| O1 | 0.0110(7) | 0.6838(10) | 0.0950(7) | 1 | 0.255(18) |
| O2 | 0.3375(6) | 0.6536(10) | 0.1009(5) | 1 | = $B_{O1}$ |
| O3 | 0.6541(8) | 0.6841(10) | 0.0904(6) | 1 | = $B_{O1}$ |
| O4 | 0.1521(8) | 0.1663(11) | 0.0919(5) | 1 | = $B_{O1}$ |
| O5 | 0.5227(7) | 0.1726(10) | 0.0906(6) | 1 | = $B_{O1}$ |
| O6 | 0.8450(7) | 0.1406(10) | 0.0813(5) | 1 | = $B_{O1}$ |

**Constrained to reflect ICP-OES data.*

**Table S3.** NPD stacking fault refinement results for $Na_{2/3}Ni_{1/3}Mn_{2/3}O_2$ in space group $P1$ with $N_V$ stacking vectors. Atomic coordinates are fixed from the co-refinement of Tables S1 and S2.

| $a$ (Å) | 8.6608(3) | | | | |
|---|---|---|---|---|---|
| $b$ (Å) | 5.01669(12) | | | | |
| $c$ (Å) | 11.1558(4)*$N_V$ | | | | |
| $R_{wp}$ | 8.82% | | | | |
| $Z$ | 12 | | | | |
| $N_V$ | 50 | | | | |
| $\mathbf{t_1}$[0,0,1] | 75% | | | | |
| $\mathbf{t_2}$[1/3, 0, 1] | 5% | | | | |
| $\mathbf{t_3}$[1/6, 1/2, 1] | 20% | | | | |
| Atom | $x$ | $y$ | $z$ | occupancy | $B_{eq}$ (Å$^2$) |
| Mn1 | 0 | 0 | 0 / $N_V$ | 1 | 0.30(2) |
| Mn2 | 0 | 0 | 0.5 / $N_V$ | 1 | = $B_{Mn1}$ |
| Mn3 | 0.5 | 0.5 | 0 / $N_V$ | 1 | = $B_{Mn1}$ |
| Mn4 | 0.5 | 0.5 | 0.5 / $N_V$ | 1 | = $B_{Mn1}$ |
| Ni1 | 0.340329 | 0 | 0 / $N_V$ | 1 | = $B_{Mn1}$ |
| Ni2 | 0.659671 | 0 | 0.5/ $N_V$ | 1 | = $B_{Mn1}$ |
| Ni3 | 0.840351 | 0.5 | 0 / $N_V$ | 1 | = $B_{Mn1}$ |
| Ni4 | 0.159649 | 0.5 | 0.5 / $N_V$ | 1 | = $B_{Mn1}$ |
| Mn5 | 0.161776 | 0.5 | 0 / $N_V$ | 1 | = $B_{Mn1}$ |
| Mn6 | 0.838224 | 0.5 | 0.5 / $N_V$ | 1 | = $B_{Mn1}$ |
| Mn7 | 0.663948 | 0 | 0 / $N_V$ | 1 | = $B_{Mn1}$ |
| Mn8 | 0.336052 | 0 | 0.5 / $N_V$ | 1 | = $B_{Mn1}$ |
| O1 | 0 | 0.686999 | 0.094664 / $N_V$ | 1 | 0.80(3) |
| O2 | 0 | 0.313001 | 0.594664 / $N_V$ | 1 | = $B_{O1}$ |
| O3 | 0 | 0.686999 | 0.405336 / $N_V$ | 1 | = $B_{O1}$ |
| O4 | 0 | 0.313001 | 0.905336 / $N_V$ | 1 | = $B_{O1}$ |
| O5 | 0.332781 | 0.651302 | 0.10193 / $N_V$ | 1 | = $B_{O1}$ |
| O6 | 0.667219 | 0.348698 | 0.60193 / $N_V$ | 1 | = $B_{O1}$ |
| O7 | 0.667219 | 0.651302 | 0.39807 / $N_V$ | 1 | = $B_{O1}$ |
| O8 | 0.332781 | 0.348698 | 0.89807 / $N_V$ | 1 | = $B_{O1}$ |
| O9 | 0.652953 | 0.684245 | 0.09079 / $N_V$ | 1 | = $B_{O1}$ |
| O10 | 0.347047 | 0.315755 | 0.59079 / $N_V$ | 1 | = $B_{O1}$ |
| O11 | 0.347047 | 0.684245 | 0.40921 / $N_V$ | 1 | = $B_{O1}$ |
| O12 | 0.652953 | 0.315755 | 0.90921 / $N_V$ | 1 | = $B_{O1}$ |

| O13 | 0.14874 | 0.166515 | 0.091935 / $N_V$ | 1 | = $B_{\mathrm{O1}}$ |
|---|---|---|---|---|---|
| O14 | 0.85126 | 0.833485 | 0.591935 / $N_V$ | 1 | = $B_{\mathrm{O1}}$ |
| O15 | 0.85126 | 0.166515 | 0.408065 / $N_V$ | 1 | = $B_{\mathrm{O1}}$ |
| O16 | 0.14874 | 0.833485 | 0.908065 / $N_V$ | 1 | = $B_{\mathrm{O1}}$ |
| O17 | 0.52603 | 0.172388 | 0.091332 / $N_V$ | 1 | = $B_{\mathrm{O1}}$ |
| O18 | 0.47397 | 0.827612 | 0.591332 / $N_V$ | 1 | = $B_{\mathrm{O1}}$ |
| O19 | 0.47397 | 0.172388 | 0.408668 / $N_V$ | 1 | = $B_{\mathrm{O1}}$ |
| O20 | 0.52603 | 0.827612 | 0.908668 / $N_V$ | 1 | = $B_{\mathrm{O1}}$ |
| O21 | 0.842193 | 0.139866 | 0.08044 / $N_V$ | 1 | = $B_{\mathrm{O1}}$ |
| O22 | 0.157807 | 0.860134 | 0.58044 / $N_V$ | 1 | = $B_{\mathrm{O1}}$ |
| O23 | 0.157807 | 0.139866 | 0.41956 / $N_V$ | 1 | = $B_{\mathrm{O1}}$ |
| O24 | 0.842193 | 0.860134 | 0.91956 / $N_V$ | 1 | = $B_{\mathrm{O1}}$ |
| Na1 | 0 | 0 | 0.25 / $N_V$ | 1 | 2.77(12) |
| Na2 | 0 | 0 | 0.75 / Nv | 1 | = $B_{\mathrm{Na1}}$ |
| Na3 | 0.691696 | 0.347186 | 0.24763 / $N_V$ | 1 | = $B_{\mathrm{Na1}}$ |
| Na4 | 0.308304 | 0.652814 | 0.74763 / $N_V$ | 1 | = $B_{\mathrm{Na1}}$ |
| Na5 | 0.308304 | 0.347186 | 0.25237 / $N_V$ | 1 | = $B_{\mathrm{Na1}}$ |
| Na6 | 0.691696 | 0.652814 | 0.75237 / $N_V$ | 1 | = $B_{\mathrm{Na1}}$ |
| Na7 | 0.5 | 0.874528 | 0.25 / $N_V$ | 1 | = $B_{\mathrm{Na1}}$ |
| Na8 | 0.5 | 0.125472 | 0.75 / $N_V$ | 1 | = $B_{\mathrm{Na1}}$ |

**Table S4.** NPD refinements of $Na_{2/3}Ni_{1/3}Mn_{2/3}O_2$ with different stacking fault probabilities for vectors $\mathbf{t}_1$, $\mathbf{t}_2$, and $\mathbf{t}_3$.

| Trial | $\mathbf{t}_1$ (%) | $\mathbf{t}_2$ (%) | $\mathbf{t}_3$ (%) | $R_{wp}$ (%) |
|---|---|---|---|---|
| 1 | 100 | 0 | 0 | 11.76 |
| 2 | 60 | 20 | 20 | 10.54 |
| 3 | 70 | 10 | 20 | 9.11 |
| 4 | 75 | 5 | 20 | 8.82 |
| 5 | 80 | 5 | 15 | 9.02 |
| 6 | 85 | 5 | 10 | 9.63 |

**Table S5.** sXRD refinement results for $Na_{2/3}Ni_{1/3}Mn_{2/3}O_2$ in space group *Cmcm*.

| | | | | | |
|---|---|---|---|---|---|
| $a$ (Å) | 2.884082(9) | | | | |
| $b$ (Å) | 5.012841(11) | | | | |
| $c$ (Å) | 11.14889(2) | | | | |
| $V$ (Å$^3$) | 161.1840(10) | | | | |
| $Z$ | 4 | | | | |
| $R_{wp}$ | 4.46% | | | | |
| Atom | $x$ | $y$ | $z$ | occupancy | $B_{eq}$ (Å$^2$) |
| Na1 | 0 | 0.6394(8) | 0.25 | 0.4109(16) | 2.43(7) |
| Na2 | 0 | 0.0413(14) | 0.25 | $= 2/3 - occ_{Na1}$ | $= B_{Na1}$ |
| Ni1 | 0 | 0 | 0 | 0.3333 | 0.670(7) |
| Mn1 | 0 | 0 | 0 | 0.6667 | $= B_{Ni1}$ |
| O1 | 0 | 0.6724(3) | 0.59055(10) | 1 | 0.76(2) |

**Table S6.** Rietveld refinements when allowing the sodium stoichiometry to freely vary.

| $x$ in $Na_xNi_{1/3}Mn_{2/3}O_2$ (Coulomb counting) | $x$ in $Na_xNi_{1/3}Mn_{2/3}O_2$ (Rietveld method) |
|---|---|
| 0.67 | 0.658(3) |
| 0.59 | 0.573(2) |
| 0.50 | 0.483(3) |
| 0.42 | 0.405(3) |
| 0.33 | 0.333(2) |

**Table S7.** sXRD refinement results for $Na_{0.59}Ni_{1/3}Mn_{2/3}O_2$ in space group *Cmcm*.

| | | | | | |
|---|---|---|---|---|---|
| $a$ (Å) | 2.875285(2) | | | | |
| $b$ (Å) | $4.980139 = \sqrt{3}a$ | | | | |
| $c$ (Å) | 11.202784(16) | | | | |
| $V$ (Å$^3$) | 160.416(0) | | | | |
| $Z$ | 4 | | | | |
| $R_{wp}$ | 3.45% | | | | |
| Atom | $x$ | $y$ | $z$ | occupancy | $B_{eq}$ (Å$^2$) |
| Na1 | 0 | 0.6667 | 0.25 | 0.3504(17) | 2.51(6) |
| Na2 | 0 | 0 | 0.25 | 0.2225(14) | $= B_{Na1}$ |
| Ni1 | 0 | 0 | 0 | 0.3333 | 0.434(4) |
| Mn1 | 0 | 0 | 0 | 0.6667 | $= B_{Ni1}$ |
| O1 | 0 | 0.6667 | 0.59045(8) | 1 | 0.690(15) |

**Table S8.** sXRD refinement results for $Na_{1/2}Ni_{1/3}Mn_{2/3}O_2$ in space group *Cmcm*.

| | | | | | |
|---|---|---|---|---|---|
| $a$ (Å) | 2.859206(11) | | | | |
| $b$ (Å) | 4.961663(17) | | | | |
| $c$ (Å) | 11.28295(3) | | | | |
| $V$ (Å$^3$) | 160.0650(10) | | | | |
| $Z$ | 4 | | | | |
| $R_{wp}$ | 4.94% | | | | |
| Atom | $x$ | $y$ | $z$ | occupancy | $B_{eq}$ (Å$^2$) |
| Na1 | 0 | 0.6488(19) | 0.25 | 0.2561(17) | 1.84(8) |
| Na2 | 0 | 0.029(2) | 0.25 | $= 1/2 - occ_{Na1}$ | $= B_{Na1}$ |
| Ni1 | 0 | 0 | 0 | 0.3333 | 0.463(7) |
| Mn1 | 0 | 0 | 0 | 0.6667 | $= B_{Ni1}$ |
| O1 | 0 | 0.6686(5) | 0.58944(11) | 1 | 0.64(2) |

**Table S9.** sXRD refinement results for $Na_{0.42}Ni_{1/3}Mn_{2/3}O_2$ in space group *Cmcm*.

| | | | | | |
|---|---|---|---|---|---|
| $a$ (Å) | 2.853639(5) | | | | |
| $b$ (Å) | 4.942648 = $\sqrt{3}a$ | | | | |
| $c$ (Å) | 11.36692(3) | | | | |
| $V$ (Å$^3$) | 160.3250(10) | | | | |
| $Z$ | 4 | | | | |
| $R_{wp}$ | 3.40% | | | | |
| Atom | $x$ | $y$ | $z$ | occupancy | $B_{eq}$ (Å$^2$) |
| Na1 | 0 | 0.6667 | 0.25 | 0.272(2) | 3.21(12) |
| Na2 | 0 | 0 | 0.25 | 0.1334(16) | = $B_{Na1}$ |
| Ni1 | 0 | 0 | 0 | 0.3333 | 0.691(6) |
| Mn1 | 0 | 0 | 0 | 0.6667 | = $B_{Ni1}$ |
| O1 | 0 | 0.6667 | 0.58868(9) | 1 | 0.63(2) |

**Table S10.** sXRD refinement results for $Na_{1/3}Ni_{1/3}Mn_{2/3}O_2$ in space group *Cmcm*.

| | | | | | |
|---|---|---|---|---|---|
| *a* (Å) | 2.84573(7) | | | | |
| *b* (Å) | 4.93177(7) | | | | |
| *c* (Å) | 11.36597(3) | | | | |
| *V* (Å$^3$) | 159.516(5) | | | | |
| *Z* | 4 | | | | |
| $R_{wp}$ | 4.85% | | | | |
| Atom | *x* | *y* | *z* | occupancy | $B_{eq}$ (Å$^2$) |
| Na1 | 0 | 0.6359(19) | 0.25 | 0.3333 | 1.44(14) |
| Na2 | 0 | 0.00(3) | 0.25 | 0 | = $B_{Na1}$ |
| Ni1 | 0 | 0 | 0 | 0.3333 | 0.882(15) |
| Mn1 | 0 | 0 | 0 | 0.6667 | = $B_{Ni1}$ |
| O1 | 0 | 0.6491(7) | 0.58929(17) | 1 | 0.17(5) |

**Table S11.** sXRD refinement results for $Na_{1/2}Ni_{1/3}Mn_{2/3}O_2$ in space group *Pnmm*.

| | | | | | |
|---|---|---|---|---|---|
| *a* (Å) | 4.961805(11) | | | | |
| *b* (Å) | 5.718390(14) | | | | |
| *c* (Å) | 11.283167(19) | | | | |
| *V* (Å$^3$) | 320.1430(10) | | | | |
| *Z* | 8 | | | | |
| $R_{wp}$ | 3.47% | | | | |
| Atom | *x* | *y* | *z* | occupancy | $B_{eq}$ (Å$^2$) |
| Na1 | 0.3514(8) | 0.5 | 0.5 | 1 | 1.92(5) |
| Na2 | –0.0318(9) | 0 | 0.5 | 1 | = $B_{Na1}$ |
| Ni1 | 0 | 0 | –0.25 | 0.3333 | 0.486(5) |
| Ni2 | 0.5 | –0.25 | –0.25 | 0.3333 | = $B_{Ni1}$ |
| Mn1 | 0 | 0 | –0.25 | 0.6667 | = $B_{Ni1}$ |
| Mn2 | 0.5 | –0.25 | –0.25 | 0.6667 | = $B_{Ni1}$ |
| O1 | 0.3267(10) | 0 | –0.1596(5) | 1 | 0.622(18) |
| O2 | 0.3354(10) | 0.5 | –0.1667(5) | 1 | = $B_{O1}$ |
| O3 | –0.1668(9) | -0.25 | –0.1581(3) | 1 | = $B_{O1}$ |

**Table S12.** Orthorhombic distortion ($\delta_o$) for $Na^+$-vacancy ordered samples with two repeats of each, refined from sXRD data.

| $x$ in $Na_xNi_{1/3}Mn_{2/3}O_2$ | $\delta_o$ Repeat 1 (%) | $\delta_o$ Repeat 2 (%) |
|---|---|---|
| 2/3 | 0.3496(4) | 0.3527(5) |
| 1/2 | 0.1893(5) | 0.1889(6) |
| 1/3 | 0.057(3) | 0.014(3) |

**Table S13.** sXRD refinement results for $Na_{2/3}Ni_{1/3}Mn_{2/3}O_2$ at 200 ºC in space group *Cmcm*.

| | | | | | |
|---|---|---|---|---|---|
| $a$ (Å) | 2.889203(9) | | | | |
| $b$ (Å) | 5.019579(16) | | | | |
| $c$ (Å) | 11.19386(3) | | | | |
| $V$ (Å$^3$) | 162.3400(10) | | | | |
| $Z$ | 4 | | | | |
| $R_{wp}$ | 8.53% | | | | |
| Atom | $x$ | $y$ | $z$ | occupancy | $B_{eq}$ (Å$^2$) |
| Na1 | 0 | 0.6640(12) | 0.25 | 0.4109 | 4.75(11) |
| Na2 | 0 | 0.031(2) | 0.25 | 0.2558 | $= B_{Na1}$ |
| Ni1 | 0 | 0 | 0 | 0.3333 | 0.870(8) |
| Mn1 | 0 | 0 | 0 | 0.6667 | $= B_{Ni1}$ |
| O1 | 0 | 0.6677(4) | 0.59242(13) | 1 | 1.35(3) |

**Table S14.** sXRD refinement results for $Na_{1/2}Ni_{1/3}Mn_{2/3}O_2$ at 35 ºC in space group *Cmcm*.

| | | | | | |
|---|---|---|---|---|---|
| *a* (Å) | 2.86011(6) | | | | |
| *b* (Å) | 4.96329(10) | | | | |
| *c* (Å) | 11.28560(10) | | | | |
| *V* (Å$^3$) | 160.206(5) | | | | |
| *Z* | 4 | | | | |
| $R_{wp}$ | 6.42% | | | | |
| Atom | *x* | *y* | *z* | occupancy | $B_{eq}$ (Å$^2$) |
| Na1 | 0 | 0.669(11) | 0.25 | 0.25 | 1.8(3) |
| Na2 | 0 | 0.020(11) | 0.25 | 0.25 | = $B_{Na1}$ |
| Ni1 | 0 | 0 | 0 | 0.3333 | 0.96(3) |
| Mn1 | 0 | 0 | 0 | 0.6667 | = $B_{Ni1}$ |
| O1 | 0 | 0.673(3) | 0.5984(4) | 1 | 1.56(11) |